\documentclass{emulateapj}

\shorttitle{\SZ X-ray Observations of the Fermi Bubbles' Edges}  
\shortauthors{Kataoka et al.}
\usepackage{times}

\def\F{{\it Fermi}-LAT }
\def\SZ{{\it Suzaku }}
\def\R{{\it ROSAT }}

\begin{document}

\title{\SZ Observations of the Diffuse X-ray Emission \\
across the Fermi Bubbles' Edges}

\author{
J.~Kataoka\altaffilmark{1,\,2}, 
M.~Tahara\altaffilmark{1}, 
T.~Totani\altaffilmark{3}, 
Y.~Sofue\altaffilmark{3}, 
{\L}.~Stawarz\altaffilmark{4,\,5},
Y.~Takahashi\altaffilmark{1}, 
Y.~Takeuchi\altaffilmark{1},\\
H.~Tsunemi\altaffilmark{6},
M.~Kimura\altaffilmark{4}, 
Y.~Takei\altaffilmark{4}, 
C.~C.~Cheung\altaffilmark{7},
Y.~Inoue\altaffilmark{8},
and T.~Nakamori\altaffilmark{9}
}
\altaffiltext{1}{Research Institute for Science and Engineering, Waseda University, 3-4-1, Okubo, Shinjuku, Tokyo 169-8555, Japan}
\altaffiltext{2}{email: \texttt{kataoka.jun@waseda.jp}}
\altaffiltext{3}{Department of Astronomy, The University of Tokyo, Bunkyo-ku, Tokyo 113-0033, Japan}
\altaffiltext{4}{Institute of Space and Astronautical Science, JAXA, 3-1-1 Yoshinodai, Chuo-ku, Sagamihara, Kanagawa 252-5210, Japan}
\altaffiltext{5}{Astronomical Observatory, Jagiellonian University, ul. Orla 171, 30-244 Krak\'ow, Poland}
\altaffiltext{6}{Department of Earth and Space Science, Osaka University, Osaka 560-0043,  Japan}
\altaffiltext{7}{Space Science Division, Naval Research Laboratory, Washington, DC 20375, USA}
\altaffiltext{8}{Kavli Institute for Particle Astrophysics and Cosmology, Stanford University, Stanford, CA 94305, USA}
\altaffiltext{9}{Department of Physics, Faculty of Science, Yamagata University, 990-8560, Japan}

\begin{abstract}
We present \SZ X-ray observations along two edge regions 
of the Fermi Bubbles, with eight $\simeq 20$
ksec pointings across the northern part of the 
North Polar Spur (NPS) surrounding the north bubble and
six across the southernmost edge of the south bubble. 
After removing compact X-ray features, diffuse X-ray emission is clearly detected
and is well reproduced by a three-component spectral model consisting of 
unabsorbed thermal emission (temperature $kT \simeq 0.1$\,keV) from the Local Bubble (LB), 
absorbed $kT \simeq 0.3$\,keV thermal emission related to the 
NPS and/or Galactic Halo (GH), and a power-law 
component at a level consistent with the cosmic X-ray background. 
The emission measure (EM) of the 0.3\,keV 
plasma decreases by $\simeq 50\%$ toward the 
inner regions of the north-east bubble, with no accompanying 
temperature change. However, such a jump in the EM is not 
clearly seen in the south bubble data.
While it is unclear if the NPS originates from a nearby supernova remnant or is related to previous activity within/around 
the Galactic Center, our \SZ observations provide evidence suggestive of the latter scenario. In 
the latter framework, the presence of a large amount of neutral 
matter absorbing the X-ray emission as well as the existence of the $kT \simeq 0.3$\,keV 
gas can be naturally interpreted as a weak shock driven by the bubbles' expansion 
in the surrounding medium, with velocity $v_{\rm exp} \sim 300$\,km\,s$^{-1}$
(corresponding to shock Mach number $\mathcal{M} \simeq 1.5$), compressing the GH gas to form the NPS feature.
We also derived an upper limit for any non-thermal X-ray emission component associated
with the bubbles and demonstrate,
that in agreement with the findings above, the non-thermal pressure and energy estimated from
a one-zone leptonic model of its broad-band spectrum, are in rough equilibrium with that 
of the surrounding thermal plasma.
\end{abstract}

\keywords{acceleration of particles --- cosmic rays --- Galaxy: center --- Galaxy: halo --- X-rays: ISM}

\section{Introduction}

The Large Area Telescope \citep[LAT;][]{atw09} onboard the {\it Fermi} satellite
is a successor to EGRET onboard the Compton Gamma-ray Observatory  \citep{har99}, with 
much improved sensitivity and resolution over a broader energy range.  
Along with the GeV detections of various $\gamma$-ray source types \citep[e.g.,][]{2FGL}, 
yet another important LAT discovery concerns the large-scale $\gamma$-ray 
structure, the so-called ``Fermi Bubbles,'' extending for about $50^{\circ}$ 
(or 8.5\,kpc) above and below the Galactic Center (GC), with a longitudinal width $\sim 40^{\circ}$ 
\citep{dob10, su10}. It has been suggested that these 
bubbles have relatively sharp edges and are symmetric with respect to the 
Galactic plane and the minor axis of the Galactic disk. The $\gamma$-ray 
emission associated with the bubbles is characterized by a hard 
spectrum with the differential $\gamma$-ray spectral photon index $\Gamma \simeq 2$; 
no significant spectral or total intensity variations have been found within the
structure, with the exception of some hints for an additional substructure consisting of 
a cocoon and jet-like features extending in the north-west and south-east directions 
\citep{su12}. The sharp edges may suggest the presence of a shock at the bubbles' 
boundaries. If however this is the case, and if the shock is the sole acceleration 
site of the $\gamma$-ray emitting particles, both the observed flat intensity profile
and the uniformly hard spectrum within the bubbles' interior extending for 
about 10\,kpc, seem rather difficult to explain.

\begin{figure*}[th]
\begin{center}
\includegraphics[angle=0,scale=0.6]{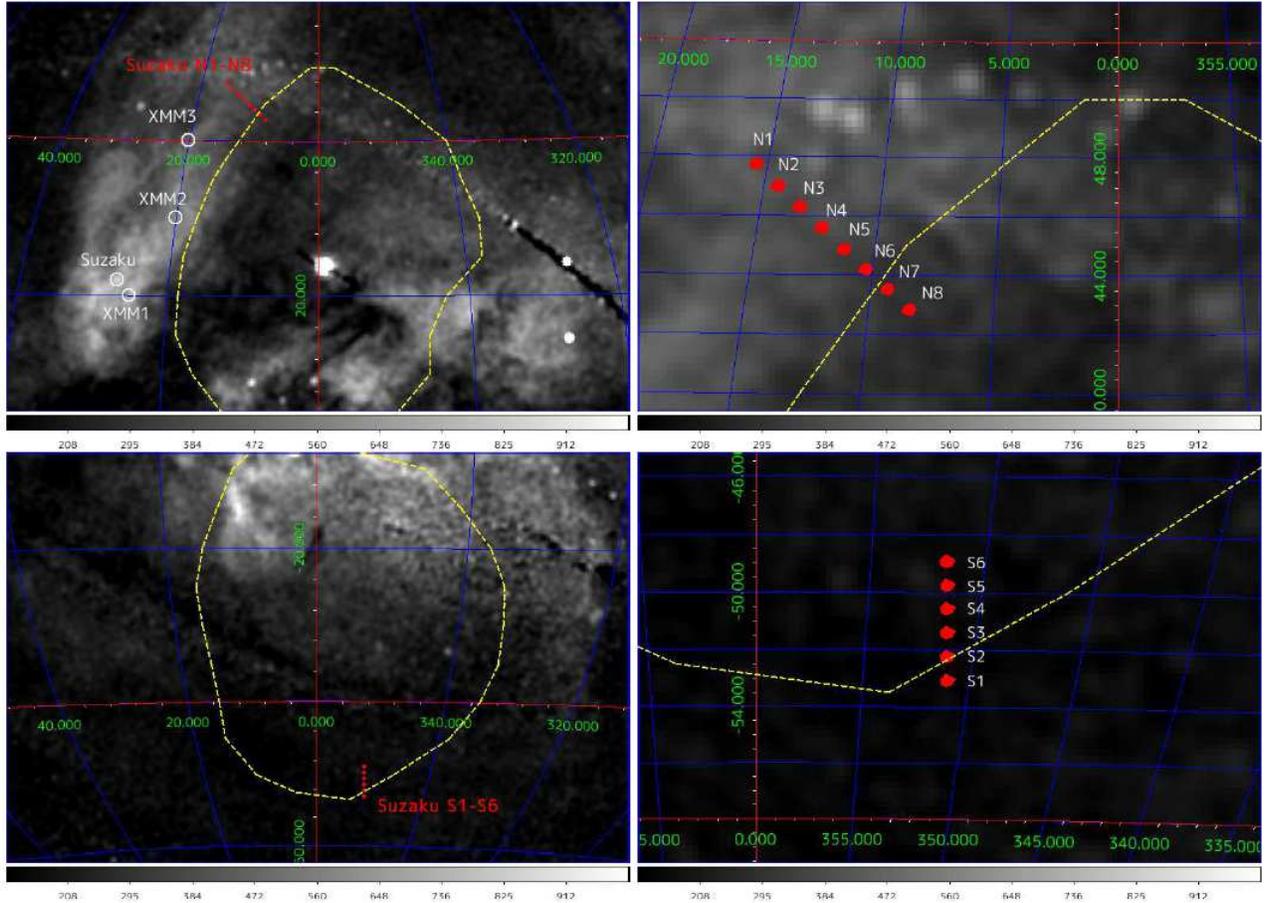}
\caption{\SZ XIS field of views ({\it red}) overlaid with the \R 0.75\,keV image for the series 
of north-east and southern observations of the Fermi Bubbles. 
All the figures are shown in Galactic coordinates. Yellow dashed lines 
indicate the boundary of the bubbles, as suggested in \cite{su10}.
{\it Top--left}: X-ray image of the north-east bubble, along with the focal centers 
of previous \SZ and {\it XMM} observations of the NPS \citep{wil03, mil08}. 
{\it Top--right}: A close-up of the north-east  edge of the bubble.
{\it Bottom--left}:  X-ray image of the southern bubble.
{\it Bottom--right}: A close-up of the southern edge of the bubble. The gray scale ranges of the \R images (units of 10$^{-6}$ cts s$^{-1}$ arcmin$^{-2}$) are indicated at the bottom of each panel.}
\end{center}
\end{figure*}

Importantly, the $\gamma$-ray emission of the bubbles is spatially correlated 
with the flat-spectrum microwave excess known as the ``WMAP haze.'' 
The haze, recently confirmed with {\it Planck} observations \citep{pla13}, is
characterized by a spherical morphology with a radius $\sim 4$\,kpc
centered at the GC, and is visible up to at least $| b |$ $\simeq$ $30^{\circ}$
\citep{fin04, dob08}. The entire microwave continuum of the haze is consistent
with a power-law distribution with spectral index $\alpha =0.5$, and is best 
interpreted as synchrotron emission from a flat-spectrum population of cosmic ray 
(CR) electrons. Hence it has been argued that the WMAP haze and the 
Fermi bubbles are linked, and that the $\gamma$-ray emission of the bubbles
is simply the inverse-Compton scattering of the cosmic microwave background 
(IC/CMB) emission from the same electron population 
producing the observed microwave synchrotron excess \citep[e.g.,][]{su10,mer11}. 
In this model, the smaller 
latitudinal extension of the WMAP haze may be due to a decrease in the magnetic 
field intensity with increasing latitude. However, the expected polarization of the haze
was not detected in the most recent accumulation of the WMAP-7 year data 
\citep{gol11}, indicating that the magnetic field in the bubbles is heavily tangled, 
or an alternative emission mechanism is at work. In this context, some authors 
argued that the bubbles/haze structure may be explained as rather being due 
to a population of relic CR protons and heavier ions injected about 
$\simeq 10$\,Gyr years ago during the outburst of a high areal density 
star-forming activity around the GC \citep{cro11}.

\begin{deluxetable*}{lllrrrrr}[th]
\tablecaption{\SZ observation log}
\tablewidth{0pt}
\tablehead{
\colhead{ID} & \colhead{start time} & \colhead{stop time} & \colhead{R.A.} & \colhead{DEC} & \colhead{$l$ } & \colhead{$b$} &  \colhead{Exposure}\\
\colhead{} & \colhead{(UT in 2012)} & \colhead{(UT in 2012)} & \colhead{[$^{\circ}$]$^a$} & \colhead{[$^{\circ}$]$^b$} & \colhead{[$^{\circ}$]$^c$} & \colhead{[$^{\circ}$]$^d$} &  \colhead{[ksec]$^e$}
}
\startdata
\multicolumn{8}{c}{Bubble North} \\
\tableline
    N1  & Aug 08  10:23 &  Aug 08 23:03  & 233.401 & 9.076 &15.480 & 47.714  & 20.0 \\  
    N2  & Aug 07  23:41 &  Aug 08 10:22  & 233.623 & 8.079 &14.388 & 47.011 & 18.3 \\  
    N3  & Aug 07  10:31 &  Aug 07 23:40  & 233.834 & 7.087 &13.321 & 46.308 & 20.4 \\  
    N4  & Aug 06  23:20 &  Aug 07 10:30  & 234.034 & 6.098 &12.280 & 45.606 & 19.0 \\  
    N5  & Aug 05  23:04 &  Aug 06 09:33  & 234.250 & 5.090 &11.255 & 44.871 & 17.2 \\  
    N6  & Aug 06  09:34 &  Aug 06 23:18  & 234.405 & 4.131 & 10.263  & 44.204 & 22.0 \\  
    N7  & Aug 08  23:06 &  Aug 09 10:20  & 234.551 & 3.174 & 9.291  & 43.537 & 19.4 \\  
    N8  & Aug 09  10:21 &  Aug 09 23:53  & 234.713 & 2.200 & 8.334  & 42.838 & 13.5 \\
\tableline
\multicolumn{8}{c}{Bubble South} \\
\tableline
    S1  & Apr 19  14:11 & Apr 20 02:44 & 332.668 & -46.192 & 351.010 &  -53.100  & 20.9 \\
    S2  & Apr 19  03:15 & Apr 19 14:10 & 331.474 & -46.348 & 351.149 &   -52.265 & 13.2 \\        
    S3  & Apr 18  04:59 & Apr 18 16:10 & 330.278 & -46.492 & 351.281 &   -51.432 & 13.5 \\        
    S4  & Apr 17  16:40 & Apr 18 04:58 & 329.080 & -46.624 & 351.406 &   -50.602 & 24.7 \\        
    S5  & Apr 18  16:12 & Apr 19 03:12 & 327.882 & -46.743 & 351.525 &   -49.775 & 21.0 \\        
    S6  & Apr 20  02:47 & Apr 20 14:25 & 326.683 & -46.851 & 351.638 &   -48.950 & 12.4 \\        
\tableline
\enddata 
\tablecomments{$^a$: Right ascension of \SZ pointing center in J2000 equinox.\\
$^b$: Declination of \SZ pointing center in J2000 equinox.\\
$^c$: Galactic longitude of \SZ pointing center.\\
$^d$: Galactic latitude of \SZ pointing center.\\
$^e$: \SZ XIS exposure in ksec.}
\end{deluxetable*}

At X-ray frequencies, the \R all-sky survey (RASS) provides full-sky images with FWHM $12'$ at 
photon energies in the $0.5 - 2$\,keV range \citep{sno95}. Hints 
of signatures of the entire north bubble structure can be noted in the 
\R map \citep[e.g.,][Figs.~18 \& 19 therein]{su10}, as well as two sharp edges in the south that trace the Fermi 
bubble below the Galactic disk (e.g., Fig.~1 in Wang 2002; Fig.~6 in Bland-Hawthorn \& Cohen 2003; also, 
Figs.~18 \& 19 in Su et al. 2010). 
Particularly noteworthy is a prominent Galactic feature called the North Polar Spur (NPS) seen in the 
0.75\,keV ($R45$ band) map, a large region of enhanced soft X-ray
emission projected above the plane of the Galaxy which coincides with a part of the radio 
Loop\,I structure. \citet[and references therein]{sof00} interpreted the NPS as a large-scale 
outflow from the GC with a corresponding total energy of $\sim 10^{55} - 
10^{56}$\,erg released on a timescale of $\sim 10^7$\,yr. This idea 
was questioned by several authors who argued that 
the NPS and the rest of the Loop\,I may instead be a nearby supernova 
remnant \citep{ber71} or may be related to the wind activity of the 
Scorpio-Centaurus OB association located at a distance of 170\,pc \citep{egg95}. 
While the GC scenario seems more plausible in view of the recent 
discovery of the Fermi bubbles \citep{su10},
the distance and origin of the NPS is still under debate, 
and we discuss this issue in more depth in this paper (Section 4.2).

Motivated by the substantial observational and theoretical progress 
recently made toward understanding the nature and origin of the large-scale structures 
extending above and below the GC regions, particularly with the discovery of the Fermi Bubbles, we conducted 
\SZ X-ray observations of high-Galactic latitude regions ($43^{\circ}$ $<$ 
$| b |$ $<$ $54^{\circ}$) positioned across the north-east and the 
southern-most edges of the bubbles.
In Section 2, we describe the details of the 
\SZ observations and data reduction procedures. The results of the analysis 
are given in Section 3, including the detailed spectral modeling and images 
obtained within the $0.4-10$\,keV range. Here we concentrated on the analysis 
of the diffuse emission and assumed all the relatively bright compact 
X-ray features detected at $\gtrsim$ 5$\sigma$ level are most likely background 
active galactic nuclei (AGN) or galaxies unrelated to the Galactic Halo (GH) and/or the NPS. Some information 
regarding these compact X-ray features are provided 
in the Appendix, along with the details of our further investigation concerning the 
data analysis. In Section 4 we summarize our findings and 
discuss the expansion of the bubbles and their interaction with the GH gas. 
We compared the obtained \SZ results with RASS observations, thus providing
guidance for future high-sensitivity all-sky X-ray studies with 
instruments like {\it MAXI} \citep{mat09} or {\it eROSITA} \citep{mer12}.  We 
also revisit the widely debated issue concerning the origin of 
and the distance to the NPS. We argue that the NPS is formed by 
plasma which has been compressed and heated by a weak shock driven
in the GH environment by the expanding bubbles. Finally, we 
discuss the energy and pressure balance between the thermal and non-thermal 
plasma components within the regions of interest by means of modeling the 
spectral energy distribution (SED) of the bubbles and of the surrounding 
medium as probed with the \SZ data. Throughout, unless otherwise noted, 
all statistical uncertainties reported  
in the paper are at the $1\sigma$ level.

\begin{figure*}[th]
\begin{center}
\includegraphics[angle=0,scale=0.65]{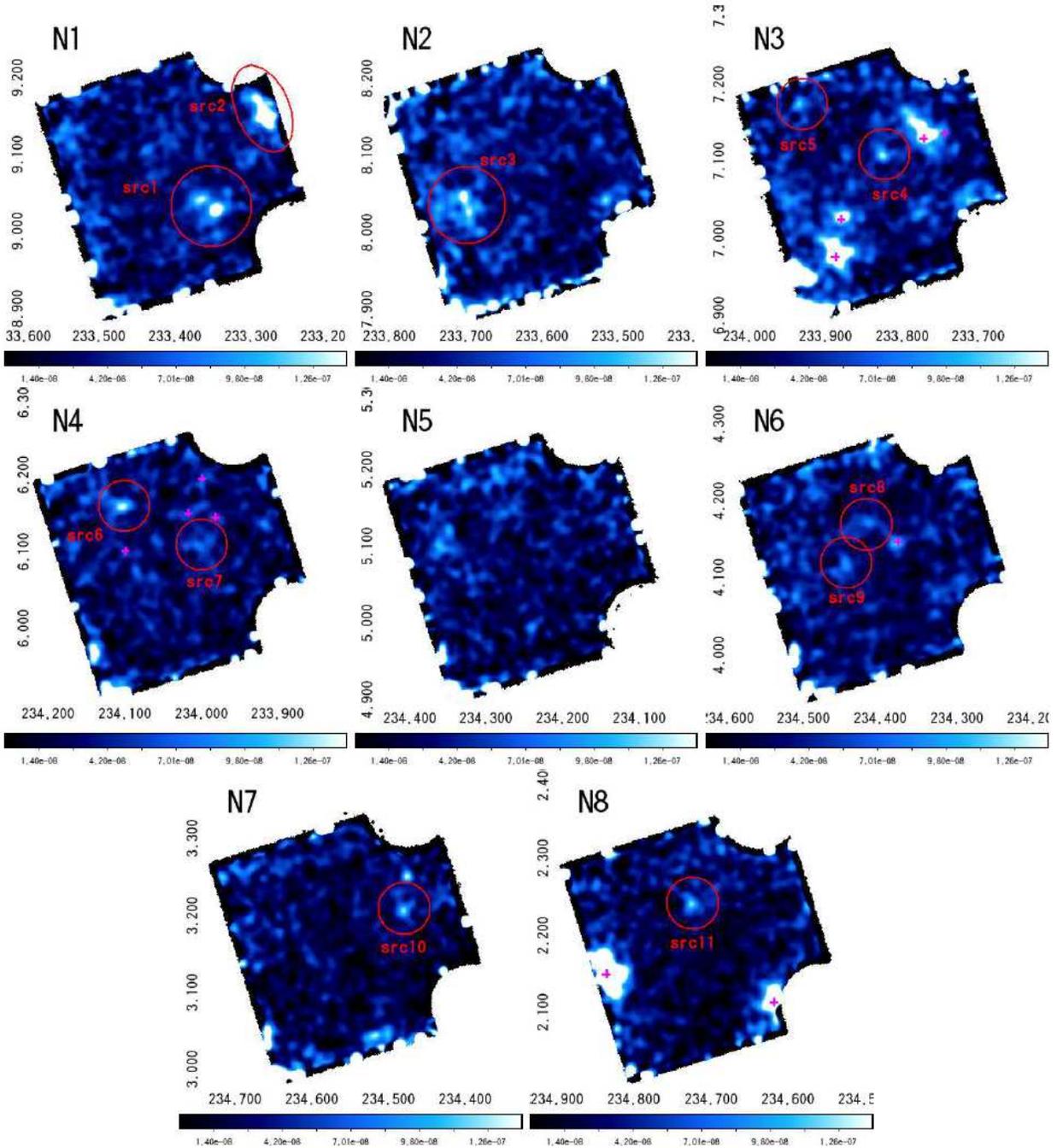}
\caption{The $0.4-10$\,keV XIS (XIS\,0+3) images of the north-east \SZ fields N1--N8,
after vignetting correction and subtraction of the ``non-X-ray background''.
Uncatalogued X-ray features detected above $\simeq 5 \sigma$ level are denoted 
as src1$-$11; the corresponding source extraction regions are denoted by red 
circles/ellipses. The size of the source extraction regions were changed 
between 1' and 2' to avoid contaminations from nearby sources. Also we 
assumed an elliptical region for src2 because it is situated at the corner of the CCD 
and hence appears elongated along the direction of the CCD edge.
The bright X-ray features that are catalogued sources are marked with 
magenta crosses. All the figures are shown in Equatorial coordinates (J2000).
\label{fig:north}}
\end{center}
\end{figure*}

\begin{figure*}[th]
\begin{center}
\includegraphics[angle=0,scale=0.65]{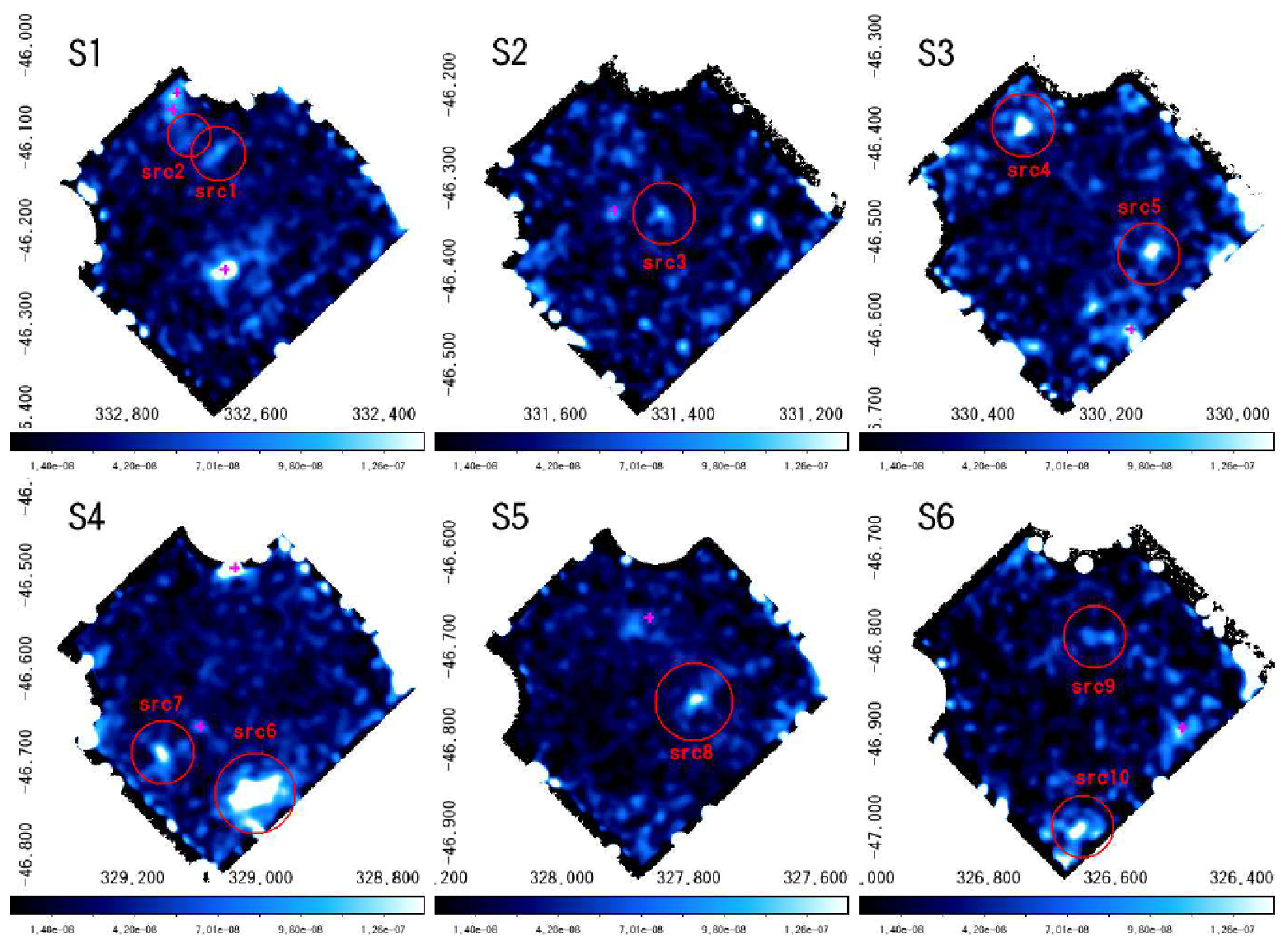}
\caption{As in figure~\ref{fig:north}, but for the southern bubble \SZ fields S1--S6.}
\end{center}
\end{figure*}

\section{Observations and Data Reduction}
\subsection{\SZ Observations}

We conducted \SZ observations of the north-east and southern edges of 
the Fermi Bubbles as a part of an AO7 program in 2012 with a
total (requested) duration of 280\,ks.  The campaign consisted 
of 14 observations, $\simeq 20$\,ksec each, consisting of eight pointings overlapping with 
the north-east bubble edge and across
part of the NPS, with the remaining six pointings over the southern 
edge of the south bubble. Table 1 summarizes the times of 
the exposures, directions of the pointing centers (in both equatorial 
and Galactic coordinates), and the effective duration of each pointing. 
The angular offset between each pointing was 1.02 deg for the north-east edge and 
0.88 deg for the south bubble edge. Because of the overlap with the NPS, we obtained 
additional pointings in the northern bubble region (spacing them out more widely) than 
in the southern bubble in order to sample a wider angular range.

The \SZ satellite \citep{mit07} carries four sets of X-ray telescopes \citep[XRT;][]{ser07},
each with a focal-plane X-ray CCD camera \citep[X-ray Imaging Spectrometer, 
XIS;][]{koy07} sensitive to photons in the energy range of $0.3-12$\,keV. 
Despite the relatively large point spread function (PSF) of the XRT with a half-power 
diameter of $\simeq 2.0'$, \SZ is the ideal instrument for the intended study because 
it provides a low and stable Non X-ray Background (NXB), particularly suitable 
for investigating extended and low surface-brightness X-ray sources \citep{mit07,taw08}. 
Three spectrometers (XIS\,0, 2, 3) have front-illuminated (FI) CCDs, while 
the XIS\,1 utilizes a back-illuminated (BI) CCD. Each XIS covers a $18' \times 18'$ 
region on the sky. Because of an anomaly in 2006 November, the operation 
of the XIS\,2 was terminated, thus here we use only the remaining 
three CCDs. The cameras were operated in the normal full-frame clocking mode 
with the $3 \times 3$ or $5 \times 5$ editing mode. Although \SZ carries also 
a hard X-ray detector \citep[HXD;][]{tak07}, hereafter we do not use the data 
collected by its PIN and GSO instruments. Figure 1 shows our \SZ XIS fields of 
view overlaid onto the \R 0.75 keV image for the series of north-east and 
southern observations of the Fermi Bubbles. Yellow dashed curves indicate 
the boundary of the Fermi Bubbles, as proposed in \cite{su10}. 
Note that at lower northern Galactic latitudes, the relatively bright NPS regions have already 
been targeted with \SZ and {\it XMM} \citep{wil03, mil08} and the corresponding 
focal centers of these archival observations are denoted in the figure by white 
circles.

\subsection{XIS Analysis}

For the XIS, we analyzed the screened data reduced using the \SZ software 
version 1.2. The reduction followed the prescriptions given in {\it `The \SZ Data 
Reduction Guide'} (also known as {\it `The ABC Guide'}) provided by the \SZ guest 
observer facility at NASA/GSFC.\footnote{http://suzaku.gsfc.nasa.gov/docs/suzaku/analysis/abc , 
http://www.astro.isas.jaxa.jp/suzaku/analysis.} 
The screening was based on the following criteria: (1) only ASCA-grade 0, 2, 3, 4, 6 
events were accumulated with hot and flickering pixels removed from the 
XIS image using the \textsc{cleansis} script \citep{day98}; (2) the time interval after 
the passage of the South Atlantic Anomaly (T\_SAA\_HXD) greater than 500\,s; 
(3) targets were located at least $5^\circ$ and $20^\circ$ above the rim of the 
Earth (ELV) during night and day, respectively.  In addition, we also selected 
the data with a cutoff rigidity (COR) larger than 6\,GV. After this screening, the 
sum of the net exposures for good time intervals was 149.8\,ks for the N1$-$N8 
and 105.7\,ks for the S1$-$S6 pointings (see Table 1). In the reduction and the analysis of 
the \SZ XIS data, the \textsc{HEADAS} software version 6.12 and the calibration 
databases (\textsc{CALDB}) released on 2012 July 11 were used, all kindly provided 
by the XIS instrumental team. The XIS cleaned event dataset was obtained in the 
combined $3 \times 3$ and $5 \times 5$ edit modes using \textsc{xselect}.

We extracted the XIS images from only the two FI CCDs (XIS 0, XIS 3) because 
the BI CCD (XIS1) has lower imaging quality due to its higher instrumental background.
The XIS images were made within three photon energy ranges of $0.4-10$\,keV, 
$0.4-2$\,keV, and $2-10$\,keV. In the image analysis, we excluded calibration sources 
at the corners of the CCD chips. The images of the NXB were obtained from 
the night Earth data using \textsc{xisnxbgen} \citep{taw08}. Since the exposure 
times for the original data were different from that of the NXB, we calculated the 
appropriate exposure-corrected original and NXB maps using \textsc{xisexpmapgen} 
\citep{ish07}. The corrected NXB images were next subtracted from the corrected 
original images. In addition, we simulated flat sky images using \textsc{xissim} 
\citep{ish07} and applied a vignetting correction. All the images obtained with 
XIS0 and XIS3 were combined. Throughout the 
reduction process, we also performed vignetting correction for all the XIS images described 
in Section 3. Finally, the images were smoothed with a Gaussian function with 
$\sigma$ = 0.07' (Gaussian kernel radius set as 8 in \textsc{ds9}).

With the combined XIS0+3 images, we first ran the source detection algorithm in 
\textsc{ximage} to discriminate compact X-ray features from 
intrinsically diffuse X-ray emission. 
This is a well established approach 
as described in \citet{gio92} and detailed in NASA's HEASARC 
Software page\footnote{https://heasarc.gsfc.nasa.gov/docs/xanadu/ximage/examples/srcdet.html. See also https://heasarc.gsfc.nasa.gov/docs/xanadu/ximage/examples/backgd.html for background calculation examples. 
In summary, the process was composed of three steps. 
First, using the \textsc{Background} command, 
we estimated the background by dividing 
an image into equal boxes characterized by the typical PSF size,  
and rejecting those not complying with certain statistical criteria. 
The average in the remaining boxes 
was taken as the background value. Second, 
with the \textsc{Excess} command, a sliding-cell 
method was used to find areas with excesses over the background threshold 
(typically, $>$3$-5$ $\sigma$). Finally, in the third step performed with the \textsc{Search} command, 
we merged the excess cells into source boxes using the \SZ PSF and 
vignetting to estimate source significances and statistics. Here we set the box sizes 
as $1.1' \times 1.1'$ (corresponding  to 64$\times$64 pixels) or $0.5' \times 0.5'$
(corresponding to 32$\times$32 pixels) for crowded fields because 
of the blurring due to the \SZ XIS PSF, and the detection threshold at 
$3 \sigma$. We performed the source detection separately in the $0.4 -10$\,keV, 
$0.4-2$\,keV, and 2$-$10\,keV ranges, and concatenated the results. }
A few spots were seen only in the $0.4-2$\,keV or 2$-$10\,keV bands, but 
these were faint sources marginally detected below the $5 \sigma$ level.  
The list and derived spectra of the bright compact emitters thus detected (all at 
approximately $\gtrsim 5 \sigma$), which are not associated 
with catalogued sources, are summarized in Appendix A. 

For the spectral analysis, we used all the FI and BI CCDs, namely, 
XIS0, 1, 3 to maximize the photon statistics.  For the bright compact X-ray features  
described above, we set the source regions to within $1'-2'$ circles
centered on the emission peaks, and estimated the backgrounds from 
annuli on the corresponding CCD chips with inner and outer radii of $2'$ and $4'$, 
respectively. The response (RMF) and auxiliary (ARF) files were produced 
using the analysis tools \textsc{xisrmfgen} and \textsc{xissimarfgen} \citep{ish07}, 
which are included in the software package HEAsoft version 6.12.  
For the  analysis of the diffuse emission, we considered the entire CCD 
chip as a source region, but only after removing all the compact features 
detected above the $3 \sigma$ level, with typical 2$'$ radius circles (see Appendix A and 
Tables 3 \& 4). The ARF files 
were produced assuming that the diffuse emission is distributed uniformly within 
circular regions with $20'$ radii (giving the ARF area of 0.35\,deg$^2$) using 
\textsc{xissimarfgen} and new contamination files (released on July 11, 2012). 
In contrast to the point source analysis, we did \emph{not} subtract background 
photons from the region in the same CCD chip but instead modeled the 
the isotropic background continuum (see Section 3.2.2 for details).

We note that even though the most recent version of \textsc{xissimarfgen} 
reproduces well the degradation of the CCD quantum efficiency below 2\,keV 
due to contamination\footnote{\texttt{http://heasarc.gsfc.nasa.gov/docs/astroe/prop\_tools/suzaku\_td/node10.html}}, 
significant differences in the spectral shape of the measured emission continua 
between XIS\,0, 1, and 3 were found below $\simeq 0.5$\,keV. Therefore we did 
not use the spectral data below 0.6\,keV for XIS\,0 and 3, and below 
0.4\,keV for XIS\,1. We also did not use the high energy data above 8\,keV for 
XIS\,1 due to low photon statistics. Light curves were also constructed for 
bright X-ray compact features, none of which indicated any significant 
variability within the short exposures performed.

\section{Results}
\subsection{X-ray Images}

The obtained $0.4-10$\,keV XIS images (combined XIS\,0 and 3) are shown 
in Figure 2 for the north-east field (N1$-$N8; from {\it top-left} to {\it bottom-right}), and 
in Figure 3 for the southern field (S1$-$S6; from {\it top-left} to {\it bottom-right}). 
At first glance, one can note some difference between the different pointings
in the north, such that the surface brightness of the targeted regions gradually decreases 
across the bubble's edge toward the GC 
from the outermost (N1) to the innermost (N8) pointings.
On the other hand, this trend is not as clear in the south (from S1 to S6). In the figures,
magenta crosses mark the positions of compact X-ray features detected at $\gtrsim 5 \sigma$ 
level which are likely associated with catalogued background sources 
from the NED or SIMBAD database. The other X-ray features detected with \SZ at high 
significances, which are missing any obvious Galactic or extragalactic identifications, 
are denoted as ``src1-11'' in the north, and ``src1-10'' in the south. A few bright features
seen in the images but not listed here are spurious sources arising from instrumental
artifacts (e.g., near the corners of the CCD chips). Red circles or ellipses in Figures 2 and 3
indicate the source extraction regions for the \emph{unassociated} compact X-ray emitters.

With the given photon statistics and the relatively large PSF of the XIS, 
it is impossible to determine if the detected compact features are point-like, 
or if they are (marginally) extended. In this context, it is 
important to estimate the number of background AGN likely to be
found within each XIS field of view (FOV). Following \citet{sta13}, who utilized
the results of the {\it Chandra} Deep Field-South observations \citep{toz01}, one should 
expect the number of AGN with $0.5-2$\,keV or $2-10$\,keV fluxes $> 5 \times 
10^{-14}$\,erg\,cm$^{-2}$\,s$^{-1}$ to be $\sim 40/$deg$^2$ at most. This implies 
two or three uncatalogued AGN within each XIS FoV, which is close to the 
number of detected compact X-ray features without any obvious optical 
identification. In order to elaborate more on the possible AGN associations, 
we have investigated the available mid-infrared (MIR) maps of the areas covered by 
our \SZ exposures using the NASA Widefield Infrared Survey 
Explorer \citep[{\it WISE};][]{wri10} satellite data. The all-sky {\it WISE} data release 
was searched for sources present within radii of $1.5''$ around the positions 
of the region centers listed in the Appendix A and no infrared counterparts were found.

\begin{figure*}[th]
\begin{center}
\vspace{-0.4in}
\includegraphics[angle=0,scale=0.6]{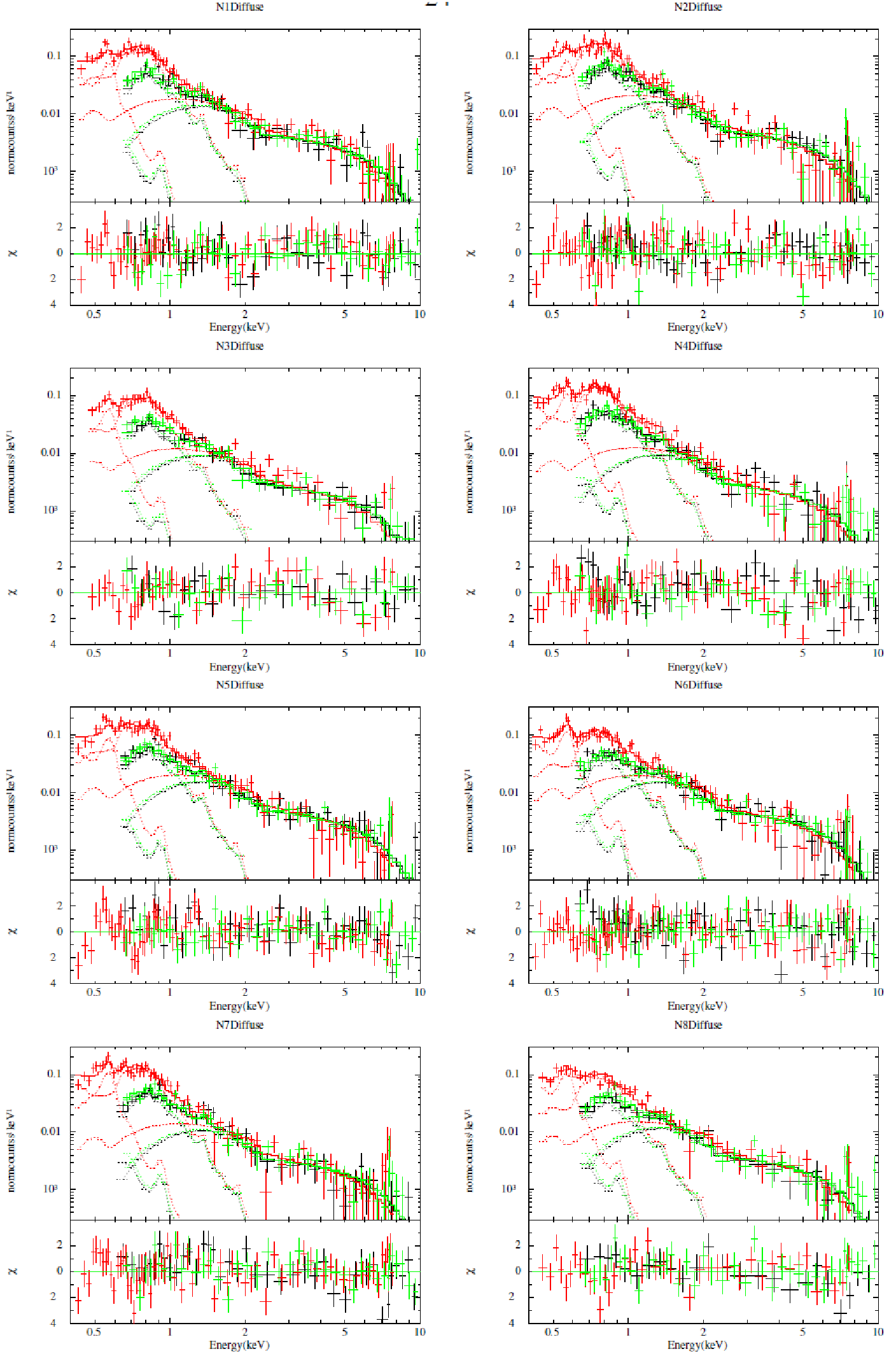}
\caption{The XIS spectra of the diffuse emission component for 
the N1$-$N8  pointings, together with the best fit model curves  
(\textsc{apec1+wabs*(apec2+pl)}) and residuals. The LB component 
($kT \simeq 0.1$\,keV; \textsc{apec1}) 
dominates the lowest 0.4$-$0.6 keV range, thermal emission related 
to the NPS and/or Galactic halo ($kT \simeq 0.3$\,keV; \textsc{apec2}) 
dominates the 0.6$-$1.5 keV range, and a power-law component from the CXB dominates 
above 1.5 keV ($\Gamma \simeq 1.4$; \textsc{pl}). 
XIS\,0 data/fits are shown in black, XIS\,1 in red, and XIS\,3 in green.
\label{fig:northspec}}
\end{center}
\end{figure*}

\begin{figure*}[th]
\begin{center}
\includegraphics[angle=0,scale=0.6]{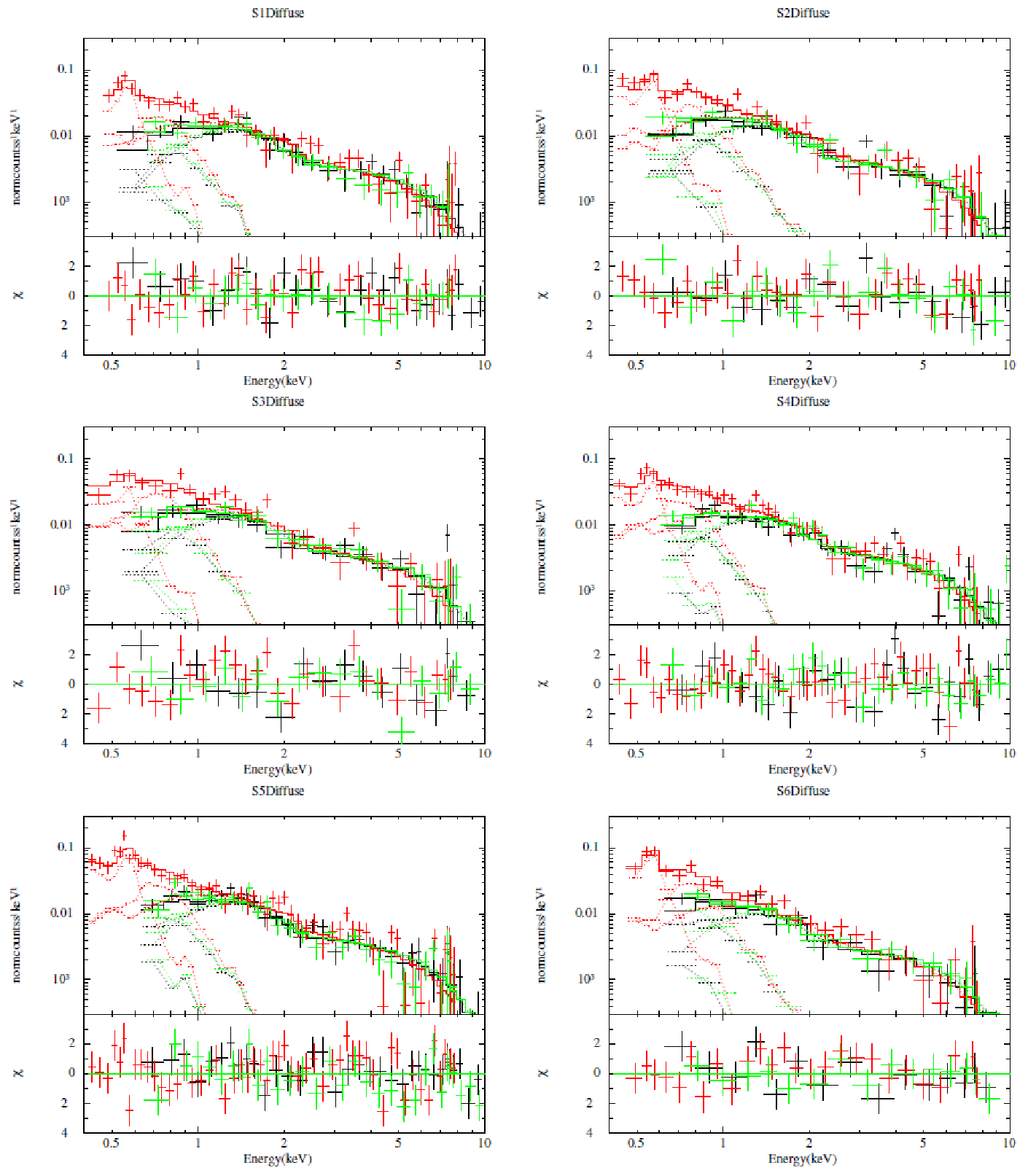}
\caption{As in figure~\ref{fig:northspec}, but for the S1$-$S6 pointings.}
\end{center}
\end{figure*}

\subsection{X-ray Spectra}
\subsubsection{Compact X-ray Features}

We first fitted the spectra of all the unassociated compact X-ray features detected 
at significance levels above $\gtrsim$ $5 \sigma$ using \textsc{xspec} (see Table 5). 
Within the given photon statistics, the X-ray spectra were all well represented 
by single power-law models moderated by the Galactic absorption, 
with the exception of src2 in the N1 field which required a substantial 
excess absorption in the neutral hydrogen column density 
$N_{\rm H} = 6.9^{+6.2}_{-3.4} \times 10^{22}$\,cm$^{-2}$ with respect to 
the Galactic value of $N_{\rm H,\,Gal}  = 3.37 \times 10^{20}$\,cm$^{-2}$.
While the nature of this source is unclear, considering only one such source detected 
in all the \SZ fields, it could be a background type-II AGN where the soft 
X-ray emission is heavily obscured by thick surrounding material \citep[e.g.,][]{kru08}.
The results of the spectral fitting for these uncatalogued X-ray emitters are tabulated 
in Appendix A. Because of the large 
spread in their fitted spectral parameters -- unabsorbed 2$-$10\,keV fluxes with a range 
$(0.1 - 7.5) \times 10^{-13}$\,erg$^{-1}$\,cm$^{-2}$\,s$^{-1}$ and 
photon spectral indices $\Gamma$ ranging from 0.8 to 3.1 -- it is difficult to draw robust 
conclusions on the nature of the compact X-ray sources. Nevertheless, the number density 
of the detected X-ray spots (two at most per individual XIS FoV) is comparable to that 
expected for background AGN as noted above (Section 3.1). Moreover, 
there is no systematic difference in the number of point-like sources detected in different 
fields at different distances from the GC. This again suggests that the compact X-ray 
features are most likely unrelated to the diffuse GH/NPS structure, but are rather 
background objects.  

\begin{deluxetable*}{ccccccccc}[th]
\tabletypesize{\scriptsize}
\tablecaption{Fitting parameters for the diffuse emission component}
\tablewidth{0pt}
\tablehead{
\colhead{ID} &  \colhead{$N_{\rm H, Gal}^a$} &  \colhead{$N_{\rm H}/N_{\rm H, Gal}^b$} & \colhead{$kT_1^c$} & \colhead{EM$_1^d$} &  \colhead{$kT_2^e$} & \colhead{EM$_2^f$} & \colhead{CXB} & \colhead{$\chi^2$/dof}\\
\colhead{} &  \colhead{($10^{20}$\,cm$^{-2}$)} &  & \colhead{(keV)} & \colhead{($10^{-2}$\,cm$^{-6}$\,pc)}  & \colhead{(keV)} & \colhead{($10^{-2}$\,cm$^{-6}$\,pc)} &   \colhead{Norm$^g$} & \colhead{}
}
\startdata
\multicolumn{8}{c}{North Bubble} \\
\tableline
N1  & 3.37 & 0.8-3.8 & 0.1(fix) & 3.47$\pm$0.38 & 0.299$^{+0.012}_{-0.010}$ & 5.49$\pm$0.38 & 1.14$_{-0.03}^{+0.04}$ & 1.07/144\\
N2  & 3.83 & 2.9-5.5 & 0.1(fix) & 3.26$\pm$0.45 & 0.315$^{+0.011}_{-0.012}$ & 5.38$\pm$0.36 & 1.03$\pm$0.03 & 1.38/162\\
N3  & 3.86 & 0.8-4.1 & 0.1(fix) & 4.09$\pm$0.71 & 0.319$^{+0.018}_{-0.015}$ & 5.28$\pm$0.50 &  1.08$\pm$0.05 &  1.07/96\\
N4  & 4.06 & 0.8-2.9 & 0.1(fix) & 4.58$\pm$0.49 & 0.324$^{+0.014}_{-0.012}$ & 5.11$\pm$0.37 &  0.71$\pm$0.04 &1.35/135\\
N5  & 4.26 & 1.1-4.8 & 0.1(fix) & 3.16$\pm$0.44 & 0.285$^{+0.010}_{-0.009}$ & 5.40$\pm$0.36 &  0.94$\pm$0.03 & 1.19/148\\
N6  & 4.45 & 0.6-4.0 & 0.1(fix) & 4.66$\pm$0.43 & 0.316$^{+0.015}_{-0.013}$ & 3.82$\pm$0.31 &  1.08$\pm$0.03 & 1.20/166\\
N7  & 4.76 & 4.4-6.7 & 0.1(fix) & 3.83$\pm$0.40 & 0.290$^{+0.011}_{-0.010}$ & 4.56$\pm$0.32 & 0.64$\pm$0.03 & 1.25/153\\
N8  & 5.02 & 1.0-5.1 & 0.1(fix) & 4.43$\pm$0.53 & 0.315$^{+0.022}_{-0.018}$ & 3.28$\pm$0.38 &  0.78$\pm$0.03 & 1.08/106\\
\tableline
\multicolumn{8}{c}{South Bubble} \\
\tableline
S1  & 1.84 & $<$ 1.8 & 0.1(fix) & 3.01$_{-0.64}^{+0.50}$ & 0.283$^{+0.063}_{-0.049}$ & 0.93$_{-0.28}^{+0.49}$ & 1.02$\pm$0.05 & 1.03/88\\
S2  & 1.66 & $<$ 2.6 & 0.1(fix) & 3.03$_{-0.55}^{+0.56}$ & 0.369$^{+0.161}_{-0.064}$ & 0.80$_{-0.28}^{+0.49}$ & 0.97$^{+0.02}_{-0.04}$ & 1.15/71\\
S3  & 1.89 & $<$ 4.1 & 0.1(fix) & 1.97$\pm$0.63 & 0.270$^{+0.077}_{-0.039}$ & 1.48$_{-0.50}^{+0.65}$ &  1.02$^{+0.04}_{-0.05}$ & 1.49/62\\
S4  & 2.16 & $<$ 2.6 & 0.1(fix) & 2.59$^{+0.51}_{-0.60}$ & 0.260$^{+0.045}_{-0.033}$ &1.37$_{-0.37}^{+0.55}$ &  1.11$\pm$0.04 &  1.15/100\\
S5  & 2.45 & $<$ 0.9 & 0.1(fix) & 3.03$^{+0.42}_{-0.52}$ & 0.262$^{+0.043}_{-0.036}$ & 1.15$_{-0.30}^{+0.50}$ & 0.92$\pm$0.03 & 1.22/117\\
S6  & 3.03 & $<$ 4.4 & 0.1(fix) & 4.06$^{+0.65}_{-0.71}$ & 0.325$^{+0.104}_{-0.051}$ & 1.18$_{-0.39}^{+0.42}$ & 0.79$\pm$0.05 & 1.22/117\\
\tableline
\enddata 
\tablecomments{
$^a$: The absorption column densities for the CXB and the GH/NPS components (\textsc{wabs*(apec2 + pl)}) components were fixed to Galactic values given in \citet{dic90}.\\
$^b$: The ratio of the absorbing column density to the full Galactic column along the line of 
sight when $N_{\rm  H}$ was left free in the spectral fitting.\\
$^c$: Temperature of the LB/SWCX plasma fitted with the \textsc{apec} model for the fixed abundance $Z = Z_{\odot}$.\\
$^d$: Emission measure of  the LB/SWCX  plasma fitted with the \textsc{apec} model for the fixed abundance $Z = Z_{\odot}$.\\
$^e$: Temperature of the GH/NPS plasma fitted with the \textsc{apec} model for the fixed abundance $Z = 0.2 \, Z_{\odot}$.\\
$^f$: Emission measure of the GH/NPS plasma fitted with the \textsc{apec} model for the fixed abundance $Z = 0.2 \, Z_{\odot}$.\\
$^g$: The normalization of the CXB in units of $5.85 \times 10^{-8}$\,erg\,cm$^{-2}$\,s$^{-1}$\,sr$^{-1}$, given in \citet{kus02} as an average of 91 observation fields, assuming a power-law model with a photon index $\Gamma_{\rm CXB} = 1.41$.}
\end{deluxetable*}

\subsubsection{Diffuse Emission}

As detailed in Section 2.2, we considered the entire CCD chip as a source region for the 
analysis of the diffuse emission
after removing all the compact features detected above $3 \sigma$ significances, 
regardless of whether they were catalogued 
or uncatalogued X-ray sources (see Appendix A and Tables 3 \& 4). 
The spectra of the \SZ pointings were all fitted
with a three component model \textsc{apec1 + wabs*(apec2 + pl)}.
The model includes an \emph{unabsorbed} thermal component (denoted as \textsc{apec1})
representing the Local Bubble (LB) emission and/or contamination from the Solar-Wind Charge 
Exchange \citep[SWCX;][]{fuj07}, an \emph{absorbed} thermal component (denoted as 
\textsc{apec2}) representing the GH or NPS, and an \emph{absorbed} power-law component 
(denoted as \textsc{pl}) corresponding to the isotropic CXB radiation. As for the absorbed 
diffuse emission, the neutral hydrogen column density was fixed to the Galactic value 
$N_{\rm H, Gal}$, in the direction of each pointing (see Table 2), while the photon 
index for the CXB component was fixed at $\Gamma_{\rm CXB} = 1.41$ \citep{kus02}. 
In the cases where $N_{\rm H}$ was left free in the 
spectral fitting, we also list the ratio of the absorbing column density to the full
Galactic column, $N_{\rm H}$/$N_{\rm H, Gal}$ (Table 2). Note that within the statistical errors,
most of the values are either consistent with unity, or even larger
than unity (for N2, N5 and N7). However, due to the relatively low photon statistics in the southern pointings, 
we can provide only upper limits hence the results are not 
conclusively close to either zero or unity.
Since the temperature and 
abundance of the LB plasma is still only poorly known, we fixed them at $kT = 0.1$\,keV 
and $Z = Z_{\odot}$, respectively, which is a common approach in the literature 
\citep{wil03,mil08, yao05, yao09, yao10}.
We note that the presence of the $kT \simeq 0.1$\,keV component related to SWCX/LB 
is confirmed by the uniform analysis of 12 and 26 fields observed with \SZ and {\it XMM}, 
respectively \citep{yos09, hen10, hen13}. 
The results of our spectral fitting are shown in Figures 4 \& 5 and are summarized in Table 2.
 
All the diffuse X-ray spectra derived for the targeted fields were well 
represented by the adopted three-component model. Below 2\,keV, the measured
continua are dominated by the $kT \simeq 0.3$\,keV thin thermal component 
(\textsc{apec2}), 
although the $kT \simeq 0.1$\,keV emission (\textsc{apec1}) is still important, especially
below 0.7\,keV, to account for the OVII (574\,eV) emission (the reduced $\chi^2$ becomes 
unacceptable if we ignore the $0.1$\,keV component attributed to LB/SWCX). Even though 
each individual spectral fit cannot precisely constrain the abundance of the 
$kT \simeq 0.3$\,keV component, we found that a sub-Solar metallicity (median $Z \simeq 
0.2 \, Z_{\odot}$) is on average preferred (see Appendix B). The depleted abundance of 
$Z < 0.5 Z_{\odot}$ is also consistent with the previously reported NPS observations with
 \SZ and {\it XMM} \citep{wil03, mil08}. Meanwhile, we do not clearly see 
the enhanced Nitrogen abundance reported in \cite{mil08} possibly due to the fact that both 
the N1$-$8 and S1$-$6 regions are much fainter than those observed in the literature and 
to our shorter exposures.
Hence, hereafter we fix the metallicity of the 
$kT  \simeq 0.3$\,keV component at $Z = 0.2 \, Z_{\odot}$. Also note that 
$N_{\rm H}$/$N_{\rm H, Gal}$ is generally large, exceeding unity for some of 
the north-east pointings, suggesting the presence of a large amount of neutral matter absorbing 
the X-ray emission of the structure.

Above 2\,keV, the X-ray spectra are dominated by the contribution of the PL
component in all the fields (N1$-$N8, S1$-$S6). Assuming this emission is isotropic, the  
2-10\,keV unabsorbed PL intensity is basically consistent with the absolute intensity of the 
CXB, namely $(5.85 \pm 0.38) \times 10^{-8}$\,erg\,cm$^{-2}$\,s$^{-1}$\,sr$^{-1}$ \citep{kus02}.
We do see some ``pointing--to--pointing'' intensity variations within the range 
$(3.8 - 6.7) \times 10^{-8}$\,erg\,cm$^{-2}$\,s$^{-1}$\,sr$^{-1}$, meaning $(0.64-1.14) \times$ 
the average CXB level (Table 2), but this level of uncertainty is naturally expected from the 
large-scale fluctuation of the CXB itself as concluded in \citet{kus02} and also 
demonstrated by 
\cite{yos09}. This indicates that we do \emph{not} see any excess non-thermal PL
emission associated with the Fermi Bubbles, at least at the level well exceeding fluctuation 
of the CXB intensity. 
To derive an upper limit for the non-thermal emission from the bubbles, we reanalyzed the 
X-ray spectra of the expected inner bubble regions (i.e., N7$-$8 and S2$-$6) by 
adding an additional PL component (with fixed photon index, $\Gamma$ = 2.0) to the one contributed by the CXB.
The obtained 90 $\%$ confidence level 
upper limit for the unabsorbed 2$-$10 keV PL emission associated with the Fermi 
Bubble is $<9.3 \times 10^{-9}$\,erg\,cm$^{-2}$\,s$^{-1}$\,sr$^{-1}$, which 
corresponds to $\sim$15$\%$ of the CXB intensity.
We revisit this issue below (Section 4.3) when modeling the overall SED of 
the Fermi Bubbles from radio to GeV $\gamma$-rays.

Finally, in Figure 6 we plot the changes in the \textsc{apec2} spectral fitting parameters, 
namely the emission measure (EM) and plasma temperature $kT$, for the 
north-east (N1$-$N8) and the southern (S1$-$S6) fields, as a function of a 
separation angle from the expected boundary of the bubble's edge \citep[following][]{su10}.
Interestingly, the EM of the $kT \simeq 0.3$\,keV component appears to change 
significantly and abruptly (by about $\simeq 50\%$) around the north-east edge 
(i.e., between pointings N6 and N7). The constant fit 
to the EM profile resulted in a $\chi^2$ value of 37.6 for 7 degree of freedom with a corresponding 
probability, $P(\chi^2) < 10^{-5}$. Note the probability 
increased significantly to $P(\chi^2)$ = 0.96 if we fit 
the EM between N1 and N5 only ($\chi^2$ = 0.61 for 4 degrees of freedom).
This change is not due to any artifact 
in the spectral fitting, e.g., an increased amount of contamination from the LB/SWXC component as we
checked in careful detail (Appendix C), although there still remains a possibility that 
we are seeing local X-ray features such as clumps or filaments $within$ the NPS rather than 
an EM change physically associated with the north-east bubble edge.
The analogous ``jump'' in the EM is not as clear in the southern bubble edge, 
and this could be partly explained in this case because we 
have only one FOV (S1) just outside the edge. But as we argue below 
(Section 4.1), since the \R 0.75\,keV map can be considered as a good tracer of 
the GH/NPS component, the absence of any NPS-like feature south from 
the GC may suggest that no sharp boundary is present there. 
Also note that the temperature of the \textsc{apec2} component is almost constant in both the north-east 
and southern fields. This temperature is a bit higher than the canonical value of 
the GH, $kT \simeq 0.2$\,keV, derived for Galactic longitudes 
$65^{\circ}$ $<$ $l$ $<$ $295^{\circ}$ \citep{yos09} or $120^{\circ}$ $<$ $l$ $<$ $240^{\circ}$ 
\citep{hen10} to avoid contamination from the GC region. Very recent measurements 
of the Galactic halo's X-ray emission using 110 {\it XMM} observations also confirm 
its temperature is fairly uniform \citep[median = 0.19 keV, interquartile range = 0.05 keV; ][]{hen13}.

\newpage

\section{Discussion}
\subsection{Comparison with RASS Maps}

In section 3.2.2, we showed that the diffuse X-ray emission, after removing compact 
X-ray sources and features detected at $\ge 3 \sigma$ level, is generally well reproduced 
by the three component plasma model: (1) $kT \simeq 0.1$\,keV emission due to 
the LB with some contamination from SWCX,  (2) $kT \simeq 0.3$\,keV plasma related to 
the GH and/or NPS gas, and (3) $\Gamma \simeq 1.4$ power-law as expected 
from the CXB. Particularly noteworthy is that the EM of the GH/NPS component 
gradually drops by $\simeq 50 \%$ around the expected boundary of the north-east Fermi 
Bubble edge. This drop is consistent with a visual impression from both \R all sky survey 
$R45$ (0.75\,keV) and $R67$ (1.5\,keV) maps, indicating that the
X-ray surface intensity gradually decreases from the N1 to N8 fields (see, Fig.~1 $top$-$left$), 
away from the NPS. However, the \R maps alone are not sufficient to quantify the
observed changes in the plasma parameters, simply because here we are dealing with a
multi-temperature convolution of various emission components and \R 
images do not provide the necessary detailed spectral information.

In order to investigate this issue in more detail, in Figure 7 ({\it top} panel) we present
the observed \R ($R45$) counting rate at 0.75\,keV, as a function of the EM (for the 
\textsc{apec2} component strictly), which we obtained from the N1$-$N8 and 
S1$-$S6 pointings, along with the EM values of the GH/NPS components 
measured at the NPS center at low Galactic latitudes in the previously reported \SZ and 
{\it XMM} observations \citep[see Figure 1;][]{wil03, mil08}. 
The horizontal error bars in the figure correspond to 
$1 \sigma$ statistical errors for the EM values estimated from the spectral model fits of the \SZ (or {\it XMM}) data (see Table 2), while 
the minimum and maximum counting rates in the \R map \citep{sno95,sno97} are given to indicate approximately 
the fluctuations present in the \R data centered on the same observing region  
\cite[e.g.,][]{yos09}. Considering the angular resolution $\sim$0.2$^{\circ}$ of the 
\R maps \citep{sno97}, 
we estimated the source counts within circles with radius $\sim$0.2$^{\circ}$. 
Here we do not exclude contributions from point features in the \R maps because no bright X-ray sources 
were in the regions of interest. In fact, we confirmed that the sum of the 
X-ray fluxes from the compact X-ray features detected by \SZ (Section 3.2.1) constitute less 
than 5 $\%$ of the diffuse emission in the 0.5$-$2 keV energy range. 
Also the contribution of unresolved sources in the 1-2 keV band measured 
by \R is 4.4$\times$10$^{-12}$ erg s$^{-1}$ cm$^{-2}$ deg$^{-2}$ \citep{has98,toz01}, 
which again corresponds to $\simeq$ 5 $\%$ of the diffuse emission in the 0.5$-$2 keV energy range.
However, we remark that even sources situated well outside the \SZ field of view (17'$\times$17') may 
contribute significantly to the reported \R counting rates, hence contamination from background sources 
may be more significant than approximated above. 
Figure 7 ({\it bottom} panel) presents the analogous counting rates measured at 
1.5\,keV ($R67$ band).

\begin{figure}
\begin{center}
\includegraphics[angle=0,scale=0.48]{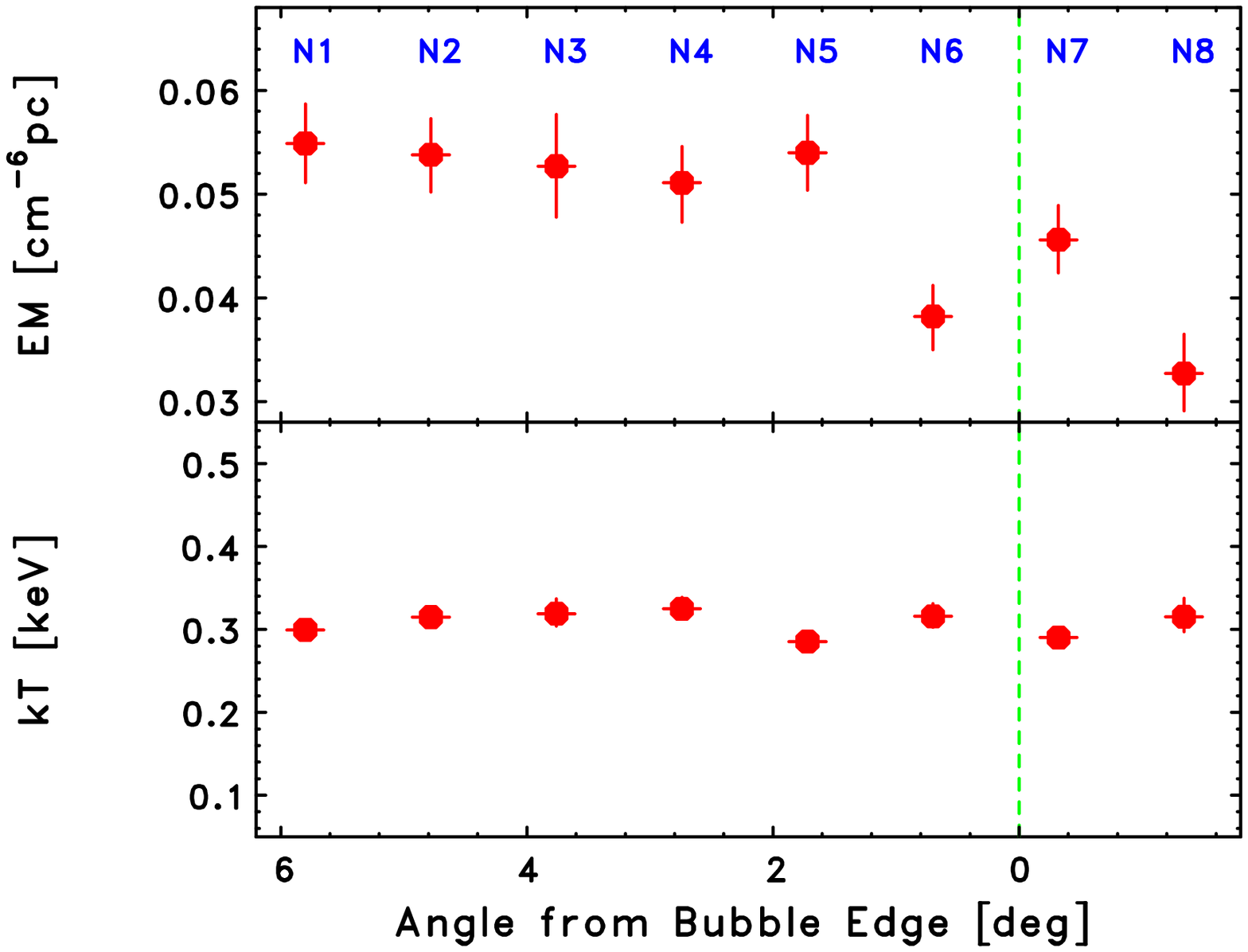}
\includegraphics[angle=0,scale=0.48]{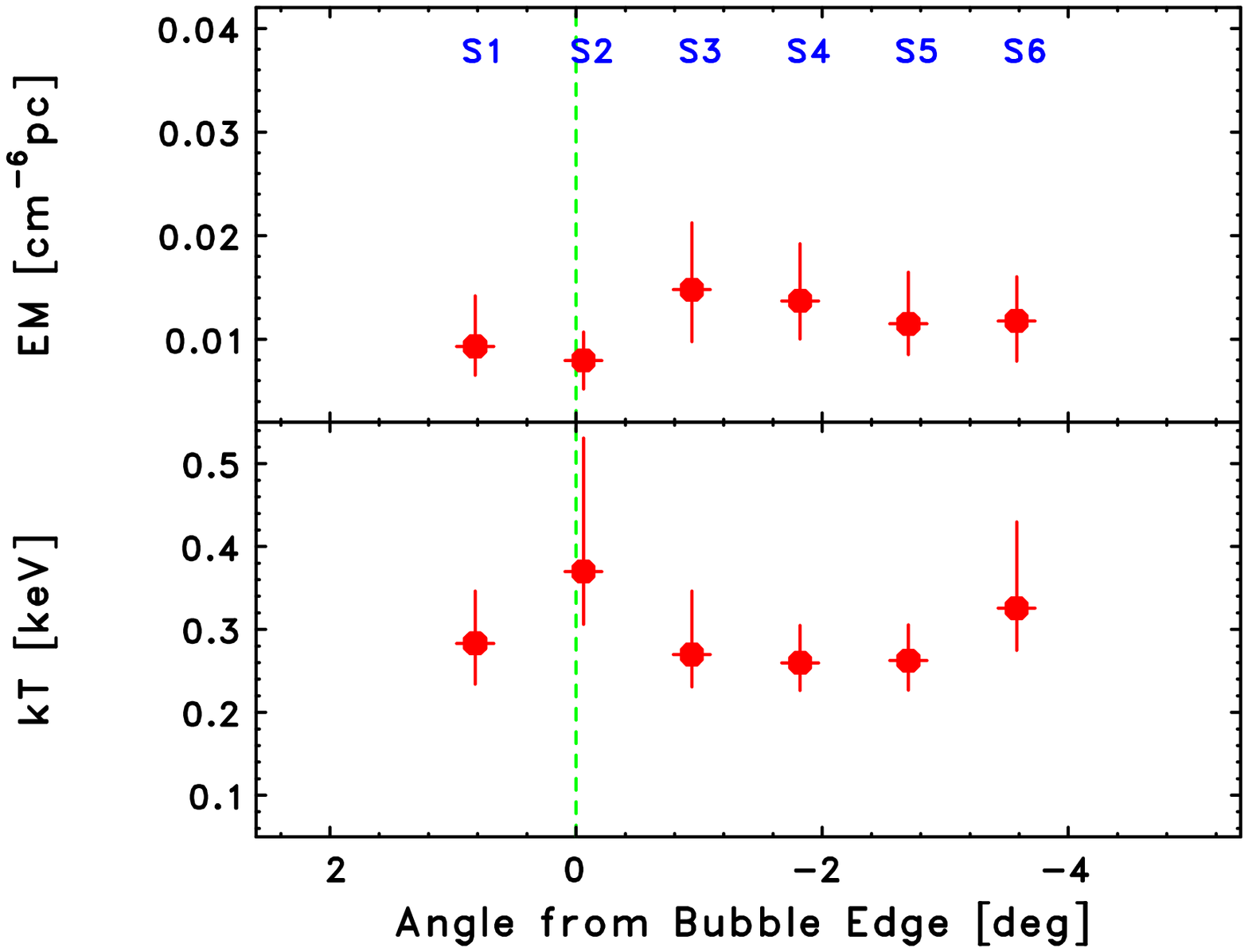}
\caption{Variation in the spectral fitting parameters EM and $kT$ for the
\textsc{apec2} emission component in the north-east ({\it top}) and south ({\it bottom}) 
bubble \SZ observations, as a function of the angular separation from the 
bubble edge. Note a significant decrease of the EM around the expected position 
of the bubbles' boundary \citep[green dotted line following][]{su10}
in the north, with no accompanying changes in the plasma temperature. 
}
\end{center}
\end{figure}

As shown in Figure 7, the $R45$ counting rates appear overall to track the EM of the GH/NPS 
component surprisingly well with a correlation factor 0.97$\pm$0.01. However, 
inspecting the data from the \SZ pointings of the north-east (N1-8) and south (S1-6) regions in more detail
suggests the correlation does not hold well within these observations. 
This could be  due to local (but small) variations in the thermal plasma parameters 
(e.g., abundance; see Appendix  C), 
differing contributions from the contaminating sources especially in the \R maps, and 
fluctuations of the CXB intensity (although this spectral component contributes negligibly 
to the $R45$ bandpass). The contribution and superposition of such background sources is 
predominantly negligible for the previous NPS pointings because they pointed at much brighter parts of the 
NPS at low Galactic latitude. Even with such complications, it appears the 
$R45$ map data can still serve as a useful overall tracer of the distribution of the GH gas even 
though the \R maps carry no explicit spectral information.
In contrast, in the $R67$ map (1.5\,keV) the contribution from the CXB becomes significant, 
diluting any intrinsic correlation between the 1.5\,keV counting rate and the EM of the 
GH/NPS component (correlation factor reduced to 0.86$\pm$0.02). In this band, additional fluctuations 
in the \R counting rates above 1 keV
could be expected due to an increased amount of background sources like AGN due to the poorer 
resolution of these data with respect to that of \SZ.

The correlation between the \R $R45$ counting rate and the $kT \simeq 0.3$\,keV 
plasma EM derived from the \SZ observations may suggest that the X-ray spectra around the 
north-east and southern Fermi Bubbles resemble each other, even though the giant 
NPS structure is seen only in the north. Indeed, what we found is that the
diffuse X-ray emission around the bubbles both above and below the GC, is well characterized 
by the three-component model introduced above, with similar values of the fitted model parameters. 
The same holds for the previously X-ray targeted fields of the bright center of the NPS analyzed 
in \citet{wil03} and \citet{mil08}. Note that in these previous works, the GH emission was 
assumed to be composed of an \emph{absorbed} $kT \simeq 0.1$\,keV component 
superimposed on the $kT \simeq 0.26$\,keV  continuum attributed to the NPS, and distinct from the GH emission. 
Although this observational fact could be a chance coincidence and we cannot rule out the possibility 
that all the emission is simply from the local structure, it appears as likely that
the $kT \lesssim 0.3$\,keV temperature gas seen at the position of the 
Fermi Bubbles, including the NPS structure, is of the same GH origin.

\subsection{NPS: Shock-Heated Galactic Halo Gas?}

Previously, it was widely agreed that the NPS and the rest of 
the Loop\,I structure arises from a recent supernova shock 
wave heating the outer shell of the superbubble, at a distance of about 
100\,pc from the Sun \citep{ber71, egg95}. Only recently, 
after the discovery of the Fermi Bubbles in particular, has the alternative interpretation 
stating that the NPS is a remnant of a starburst or a nuclear 
outburst which happened near/within the GC about 15\,Myr ago
\citep{sof77,sof84,sof94,sof00,sof03,jes03} entered back into the 
limelight. The argument against the GC scenario followed from the
measurement of the interstellar polarization at a distance of about 
100\,pc, which seems to trace some part of Loop\,I including the NPS 
\citep{bin67, mat70}. However, the observed stellar polarization orientation 
is almost \emph{perpendicular} to the direction of the NPS, especially at 
low Galactic latitudes, which is at odds with the SNR association
\citep[e.g., ][]{fur90,xu07,xia08}. 
Specifically, the NPS radio ridge at $b$ = 20$-$30$^{\circ}$ runs at angle 
130$^{\circ}$ \citep[from GC toward $l$ =90$^{\circ}$:][]{sof79}, 
while the optical polarization is at 40$-$60$^{\circ}$ \citep{mat70}. 
Thus, the implied local magnetic field orientation is nearly perpendicular to the NPS and does 
not support the local SNR origin. It has also been argued that the 
observed optical polarization is aligned with H\,I filaments of the Hydra
 ridge, which is a local HI region inflated by magnetic fields, 
and unrelated to non-thermal features like the NPS 
\citep{sof73,sof76}. The Hydra HI ridge at $b$ = 20$-$30$^{\circ}$ 
also runs at 20$-$50$^{\circ}$, nearly perpendicular 
to the NPS. Note that HI at $b$ =70$-$90$^{\circ}$ appears parallel to the NPS orientation,
and that has been taken as evidence for the HI$-$NPS association. 
However, as mentioned above, the brightest part of the NPS at $b$ =20$-$30$^{\circ}$ 
is perpendicular to HI. For further details of these arguments, see the discussion 
in \citet{sof74}.

One should note further that both the \SZ and {\it XMM} observations 
targeting any part of the NPS implied relatively 
large values of the neutral hydrogen column density $N_{\rm H}$ 
absorbing the NPS thermal spectrum in all cases. 
For example, based on the {\it XMM} data, \citet{wil03} derived 
column densities  0.9, 0.6, and 0.5 times  the Galactic value $N_{\rm H,\,Gal}$, for the
three different regions positioned at ($l$, $b$) = (25.0$^{\circ}$, 20.0$^{\circ}$), 
(20.0$^{\circ}$, 30.0$^{\circ}$), (20.0$^{\circ}$, 40.0$^{\circ}$), respectively 
(see Figure 1 for the {\it XMM} pointing positions). Similarly, \citet{mil08} 
concluded from spectral fitting of \SZ data on the 
brightest part of the NPS, ($l$, $b$) = (26.84$^{\circ}$, 21.96$^{\circ}$), 
was either $> 0.71$ or $>0.97$ times the Galactic value, depending on the choice 
of background regions. At first glance, such high levels of $N_{\rm H}$ (i.e., column 
densities more than 0.5 times the $total$ Galactic value toward in the 
line-of-sight direction), 
confirmed by our analysis of the newly acquired \SZ data, seems to conflict 
with the idea that the NPS is a local phenomenon. \cite{wil03} mention 
that the halo and NPS components lie behind at least 50$\%$ of the line-of-sight 
cold gas for which the total Galactic column density in the range 
(2$-$8)$\times$10$^{20}$ cm$^{-2}$, and attribute this high $N_{\rm H}$ to the 
cold gas distribution in the wall located at 15$-$60 pc. However, the presence of 
such a wall between the LB and NPS is not confirmed, but was rather an assumption made in order 
to not conflict with their local model. \cite{mil08} also reported high levels of 
$N_{\rm H}$ but no discussion about the origin of such large amount of 
cold gas; throughout, they assumed that the NPS is a local structure 
based on the interstellar polarization feature and the HI features, both of 
which cannot however be taken to strongly support the local interpretation as we 
discuss above.

Moreover, the inspection of the \R maps indicates that the X-ray emission 
from the NPS is heavily absorbed at 1.5\,keV at low Galactic latitudes 
\citep[e.g.,][]{sno95}, i.e., from the Galactic plane up to $b$ $\simeq$ 
10$^{\circ}$. This requires hydrogen column densities as large as 
$\gtrsim 5 \times 10^{21}$\,cm$^{-2}$. Meanwhile, any accumulation of 
neutral gas within a 100\,pc distance by an expanding shock wave should 
amount to no more than $N_{\rm H} \simeq 3 \times 10^{20}$\,cm$^{-2}$.
Therefore, it seems reasonable and natural to consider the $kT \simeq 0.3$\,keV 
plasma component detected in our \SZ observations of the north-east and southern 
Fermi Bubbles' edges is essentially the same as the plasma component 
seen in the previous observations of the NPS, having the similar 
temperature, $kT \simeq 0.3$\,keV. 
Yet it is difficult to conclude if the NPS is physically associated with the 
GH, because as already emphasized above, the derived temperature of this component 
is slightly higher than the ``canonical'' $kT \simeq 0.2$\,keV value claimed 
for the GH gas \citep{yao05, yao09, yao10, yos09,hen10, hen13}. 
On the other hand, this discrepancy can be explained as a signature of gas heating by the
expanding bubble structure, which drives a low-Mach number shock in the surrounding
medium \citep[for a high-Mach number case see, ][]{guo12a,guo12b}. 
We return to this issue in Sections 4.3 and 4.4.

Finally, let us comment in this context on the aforementioned jump in the 
EM of the hot gas component at the north-east bubble edge. Here we propose 
that the observed $50\%$ decrease in EM is likely due to projection effects 
related to a cavity inflated by an expanding bubble in the GH environment. 
Namely, assuming that the Fermi Bubbles are indeed characterized by sharp edges 
and are symmetric with respective to the observer's line of sight, a shell of
the evacuated gas is expected to form an envelope around the expanding 
structure. A projection of the emission of this shell onto the bubbles' interior
should then result in the same temperature plasma component 
(here $kT \simeq 0.3$\,keV) being observed from both within and around the bubbles
even if they are devoid of any thermal component, but only with the enhanced 
emissivity just outside the bubbles' edges. To estimate the exact shape of the 
EM profile requires detailed modeling of the emissivity profile as 
proposed to model the radial profile of shell-type supernova remnants 
\citep[e.g.,][]{ber04}, which is beyond the scope of this paper.  Instead, 
we consider a simple 2-dimensional toy model in which the uniform gas is confined 
in a donut region between $R_{\rm in}$ and $R_{\rm out}$, where  $R_{\rm in}$ 
$\simeq 4$\,kpc is the radius of the bubble and the observed EM is simply 
proportional to the path length of gas along the line of sight. Within 
such a toy model,  a $50\%$ drop of EM can be explained if the width of the 
outer shell of the bubble, $R_{\rm out} - R_{\rm in} \simeq 2$\,kpc, 
i.e., is twice smaller than the bubble radius.

\begin{figure}
\begin{center}
\includegraphics[angle=0,scale=0.5]{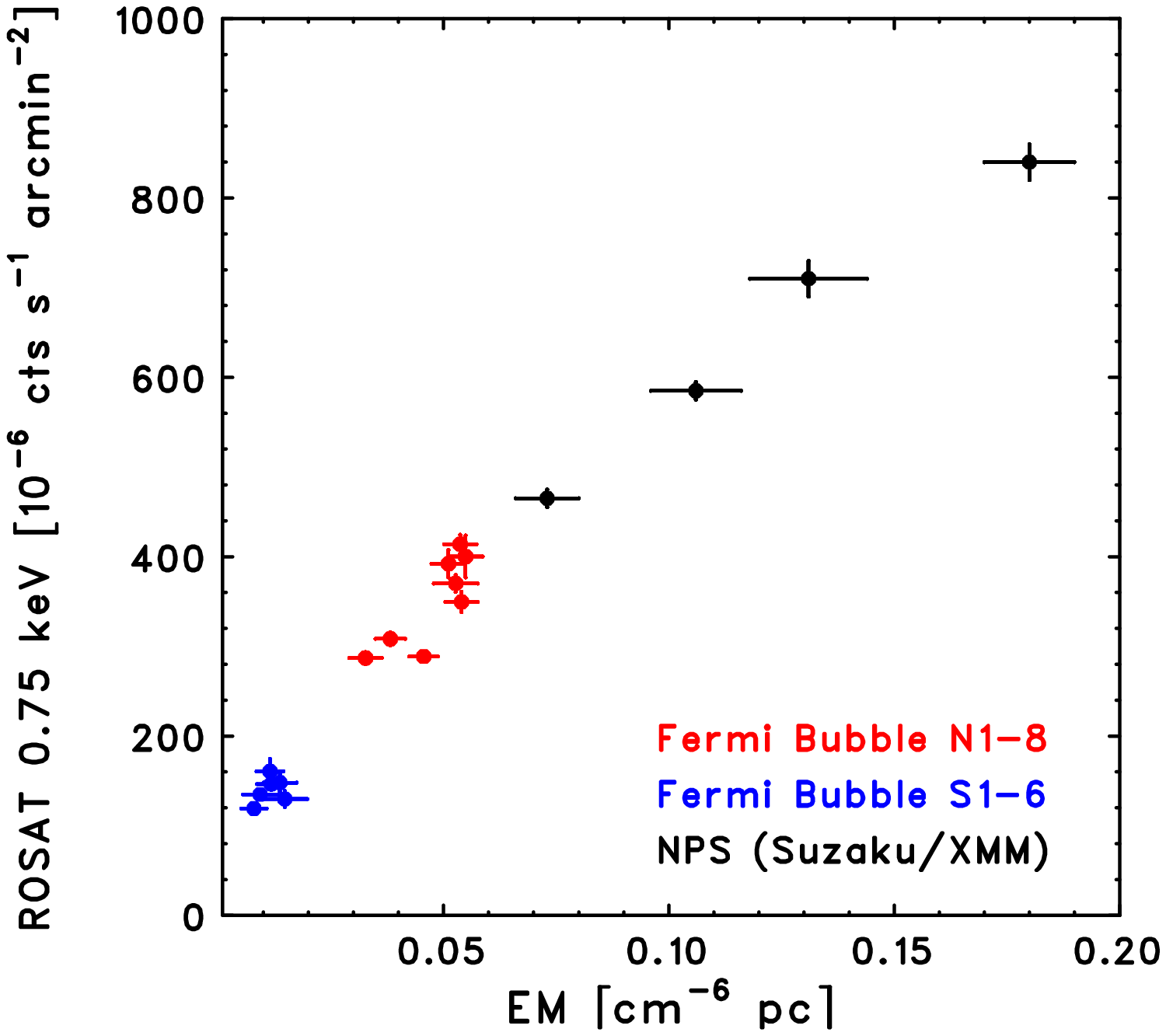}
\includegraphics[angle=0,scale=0.5]{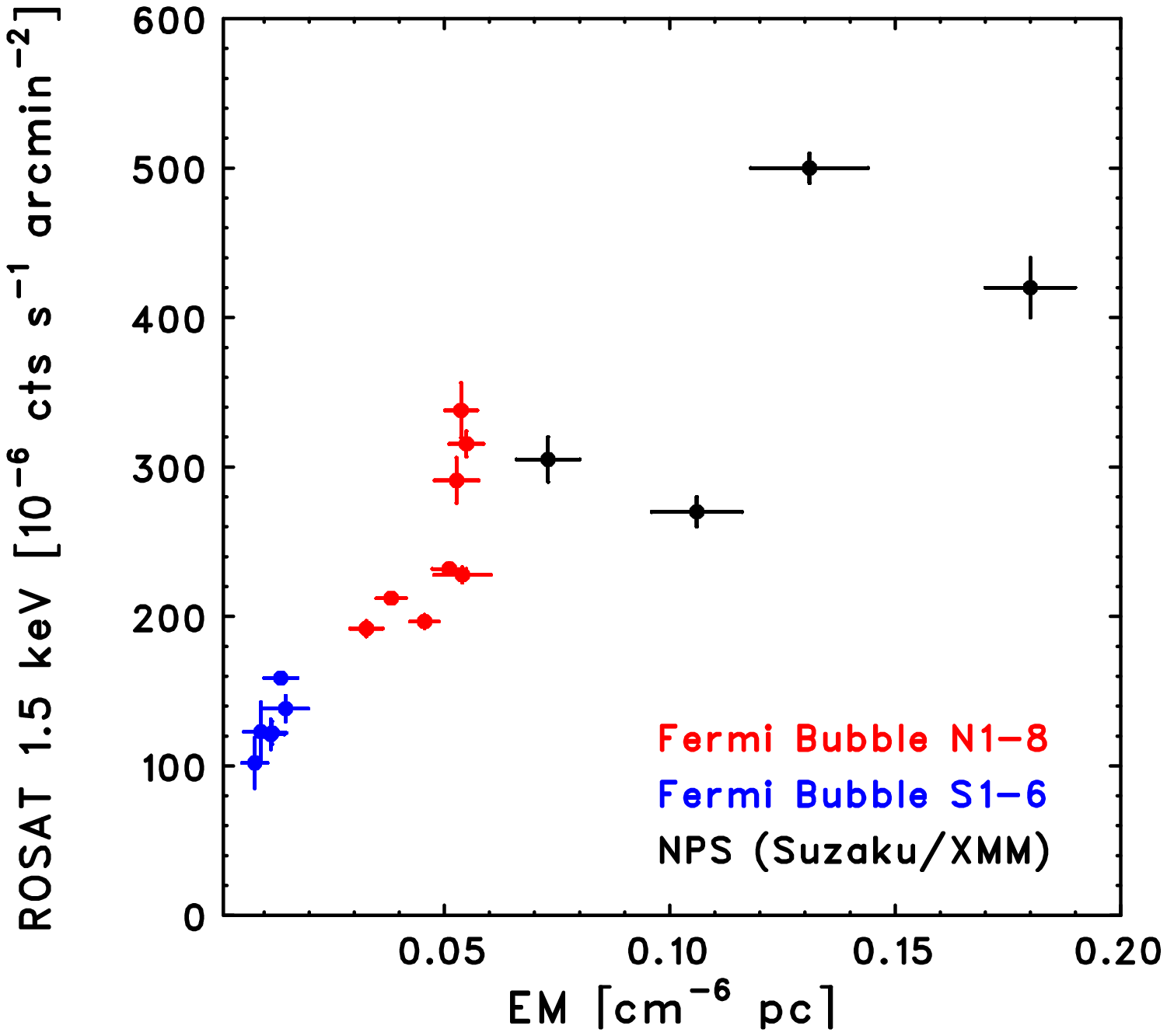}
\caption{\R count rates in units of $10^{-6}$\,cts\,s$^{-1}$, taken from the 0.75\,keV ({\it top}) 
and 1.5\,keV ({\it bottom}) maps, compared with  our \SZ estimates of the 
EM for the GH/NPS component. Included also are the archival \SZ or {\it XMM} observations 
of the NPS \citep{wil03, mil08}.}
\end{center}
\end{figure}

\subsection{Thermal vs Non-thermal Plasma}

In the spectral fitting of the newly acquired \SZ data, we did not detect 
any excess non-thermal emission associated with the bubbles, at least at 
the level exceeding the expected $\sim 10\%$ fluctuations in the CXB. 
Figure 8 shows the SED of the Fermi Bubbles, 
from radio to GeV $\gamma$-ray, with the corresponding X-ray upper limit. 
The GeV data points correspond to the emission of the entire bubbles'
structure following \citet{su10}. The radio data points corresponds to the 
WMAP haze emission averaged over $b = -20^{\circ}$ to $-30^{\circ}$, for 
$|l| < 10^{\circ}$. The bow-tie centered on the 23\,GHz $K$-band indicates 
the range of synchrotron spectral indices allowed for the WMAP haze
following \citet{dob08}. In our modeling, we assumed a simple 
one-zone leptonic model in which the radio emission and GeV $\gamma$-ray 
emission arise from the same population of relativistic electrons through
the synchrotron and inverse-Compton (IC/CMB) processes, respectively 
\citep[e.g.,][]{su10}. 
We are aware that detailed modeling requires also the IC contributions 
from the dust and starlight, i.e., far infrared and optical/UV backgrounds 
as detailed in \cite{mer11}. However, such starlight/dust emission at the position of the 
lobes is anisotropic and non-uniform, therefore special care must be taken when including these additional sources of 
seed photons. The interstellar stellar radiation field 
has energy density of $\sim$ 1 eV cm$^{-3}$, comparable to that of the CMB, 
but its contribution is more significant closer to the disk and as such, the conclusion 
is not significantly affected at high galactic latitudes. 
In fact, \citet[][Fig.~2 therein]{mer11} demonstrated that the 
IC/CMB contribution is most significant up to 10 GeV in the \F data.

\begin{figure*}[th]
\begin{center}
\includegraphics[angle=0,scale=0.7]{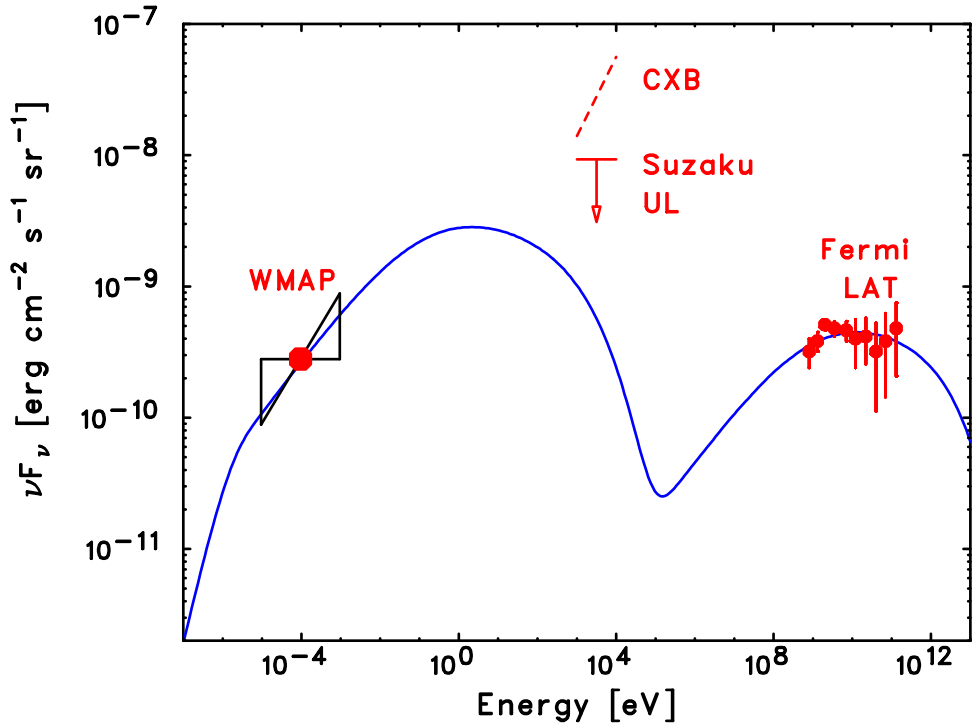}
\caption{SED of the Fermi Bubbles fitted with the one-zone leptonic model (blue curve). 
We assumed the magnetic field intensity $B = 12$\,$\mu$G within the bubbles, 
and the emission volume $V = 2 \times \frac{4}{3} \pi R^3$ with radius 
$R = 1.2 \times 10^{22}$\,cm.  Full details are given in Section 4.3.
The GeV data points correspond to the emission of the entire bubbles'
structure, following \citet{su10}. The radio data points corresponds to the 
WMAP haze emission averaged over $b = -20^{\circ}$ to $-30^{\circ}$, for 
$|l| < 10^{\circ}$. The bow-tie centered on the 23\,GHz $K$-band indicates 
the range of synchrotron spectral indices allowed for the WMAP haze,
following \citet{dob08}. Red dashed line denotes the observed CXB level, and the 
solid line indicates the \SZ upper limit for the bubbles' non-thermal X-ray 
emission, $<9.3 \times 10^{-9}$\,erg\,cm$^{-2}$\,s$^{-1}$\,sr$^{-1}$ 
in the 2$-$10 keV energy range, corresponding to $\sim$15 $\%$ of the CXB level.}
\end{center}
\end{figure*}

For the electron energy distribution we assume a standard
broken power-law form $N_e(\gamma > \gamma_{\rm min}) = N_0 \, \gamma^{-s} \, \left(1 + 
\frac{\gamma}{\gamma_{\rm brk}}\right)^{-1} \times \exp\!\left[- \frac{\gamma}{\gamma_{\rm max}}\right]$,
with the injection index $s = 2.2$, and the minimum and maximum electron Lorentz factors set 
to $\gamma_{\rm min} = 2000$ and $\gamma_{\rm max} = 10^8$, respectively. 
The parameter $\gamma_{\rm brk} = 10^6$ is the characteristic energy above 
which the electron spectrum breaks by $\Delta s = 1$. We further anticipate the 
magnetic field intensity $B = 12$\,$\mu$G within the bubbles, and the emission 
volume $V = 2 \times \frac{4}{3} \pi R^3$ with radius $R = 1.2 \times 10^{22}$\,cm.
The modeling results, shown in Figure 8 as a blue curve, yield the 
non-thermal bubbles' pressure $p_{\rm n/th} = (U_e + U_B)/3$ $\simeq 2.0 \times 
10^{-12}$\,dyn\,cm$^{-2}$, where the electron and magnetic field energy densities
are $U_e = \int d\gamma \, m_e c^2 \gamma \, N_e(\gamma)$ and $U_B = B^2/8\pi$, 
and the total non-thermal energy stored in electrons and magnetic field 
$E_{\rm n/th} = (U_e + U_B) \times V$ $\simeq 10^{56}$\,erg. The 
results of our model fitting suggest $U_B$ $\gg$ $U_e$. 
Note that the derived $B = 12$\,$\mu$G seems a factor of 
2$-$4 higher than the typical magnetic field in our Galaxy, but consistent with independent 
estimates in the literature of $B = 15$\,$\mu$G \citep{mer11} or $B = 5-10$\,$\mu$G 
\citep{su10}. Following \cite{mer11}, this could be due to an overestimate of
microwave flux caused by inappropriate template subtraction. In fact, the model prediction 
assuming  $B = 4$\,$\mu$G suggests an order of magnitude smaller flux for the 
WMAP haze so more detailed modeling of the SED will be justified only after updated
 {\it Planck} and \F measurements are made available. Similarly at this 
stage, we can neither support or rule out various other models including 
the hadronic model we mentioned in the introduction section \citep{cro11}. Therefore, 
our IC/CMB model should be considered as one possible interpretation
that should be clarified and tested in the near future.

For comparison, we can also estimate the thermal pressure of the NPS 
gas as $p_{\rm th} \simeq n_g \times kT$, where $n_g$ is the gas number density 
and $kT$ is the gas temperature. 
From our \SZ observations we take $kT \simeq 0.3$\,keV, and estimate
$n_g = (\rm{EM}/{\it d})^{1/2}$, where $d$ is the scale length 
of the X-ray plasma with the given emission measure EM.
Assuming the thickness of a thermal X-ray envelope/shell (see the discussion in
Section 4.2) as $d \simeq 2$\,kpc, we obtain $p_{\rm th} \simeq$ 
$2 \times 10^{-12}$\,dyn\,cm$^{-2}$ and $E_{\rm th} \simeq$ $10^{56}$\,erg. 
Even though all these estimates are rough, and are based on clearly over-simplified
modeling, they indicate robustly that under all the model assumptions specified 
above, the non-thermal plasma 
filling the Fermi Bubbles 
and the thermal plasma of the bubbles' immediate surroundings are in pressure and energy equipartition.
This finding is in accord with the idea that the NPS feature is
composed of the GH gas heated by a shock wave driven by the expanding bubbles.
Indeed, in such a situation, pressure equilibrium between shock downstream
fluids is expected.

In the framework of the above interpretation, the Mach number of a shock 
following from the observed temperature ratio $kT_+/kT_- \simeq 0.3\,{\rm keV}/0.2\,{\rm keV}$
is $\mathcal{M} \simeq 1.5$, assuming the adiabatic parameter of the GH gas 
$\hat{\gamma} = 5/3$. This further implies the upstream (unperturbed GH gas)
pressure $p_- \simeq$ $0.8 \times 10^{-12}$\,dyn\,cm$^{-2}$, and the shock
velocity $v_{\rm sh} \simeq \mathcal{M} \times c_{s-} \simeq 320$\,km\,s$^{-1}$
for the upstream sound speed $c_{s-} \simeq 200$\,km\,s$^{-1}$. As discussed
below, the estimated value of $v_{\rm sh}$ is in agreement with the expected
expansion velocity of the Fermi Bubbles.

\subsection{On the Formation of the Fermi Bubbles}

Let us comment here on the formation of the Fermi Bubbles in the context of the
presented \SZ observations.  Note again that the discussion below is our 
speculation based on an assumption that both the Fermi Bubbles and NPS are 
connected with the GC past activity. Therefore as we have shown above, 
the local bubble scenario for the NPS can also work in some sense and 
still leaves a lot of room to be clarified in future works.
Nevertheless, there are a number of observations discussed in the literature 
taken as evidence that the GC has undergone 
multiple past epochs of enhanced activity on different timescales, due either to 
AGN-like outbursts or episodes of circumnuclear starbursts.
The strongest case among these is the Fe K$_{\alpha}$ echo from molecular clouds 
situated a few hundreds parsec apart around 
Sgr\,A$^{\star}$ \citep{koy96,mur00,mur01}.
Recently, a diffuse clump in an over-ionized state with a jet-like structure 
has also been found in the \SZ data for the GC south, $\sim$200 pc 
from  Sgr\,A$^{\star}$, suggesting a plasma ejection from Sgr\,A$^{\star}$ 
which happened about a million years ago (Nakashima et al. 2013, in prep). 
Outflows of this kind are expected to lead to the formation of bubbles/lobes 
expanding within the Galactic halo, just like the GC scenario of NPS as well as
the Fermi Bubbles, sweeping up the interstellar/halo gas in analogy with the extended 
lobes seen in distant radio galaxies \citep{sof00}. 
\cite{tot06} has shown that the Fe K$_{\alpha}$ echo, NPS, and the
observed 511 keV line emission toward the GC can be explained naturally in a
standard framework of a radiatively inefficient accretion flow (RIAF) in the GC black hole,
if the typical accretion rate was about 1,000 times higher than the
current rate in the past 10 Myr. The outflow energy expected
by such an accretion rate is expected to be 10$^{56}$ erg
(or 3$\times$10$^{41}$ erg s$^{-1}$). 

The GH is thought to be rather isothermal, characterized by a
temperature, $kT \simeq 0.2$\,keV, 
with only some density gradients towards the GC \citep{yos09,hen10, hen13}.
During the evolution of the outflow, the evacuated halo gas can be heated, if the
bubbles' expansion is supersonic, due to formation of a shock wave at the edges of the
structure. In the previous section we estimated the Mach number
of a shock needed to heat the GH gas from $kT \simeq 0.2$\,keV up to 
$kT \simeq 0.3$\,keV characterizing the NPS emission as $\mathcal{M} \simeq 1.5$, 
corresponding essentially to the trans-sonic expansion velocity of the Fermi Bubbles 
$v_{\rm exp} \sim 300$\,km\,s$^{-1}$. Interestingly, this is comparable to the escape 
velocity from the Galaxy. If we assume the velocity is approximately constant during the 
entire evolution, the characteristic timescale for the formation of the observed structure is 
$t_{\rm exp} \simeq R/v_{\rm exp}$ $\simeq 10$\,Myr, where $R \simeq 4$\,kpc 
is the radius of the bubbles. With the total non-thermal energy estimated above, 
$E_{\rm n/th} \simeq$ $10^{56}$\,erg, the required time-averaged jet/outflow 
kinetic luminosity then reads
as $L_{\rm jet} \simeq E_{\rm n/th}/t_{\rm exp}$ $\simeq 3 \times 10^{41}$\,erg\,$s^{-1}$,
which is $\sim 0.1\%$ of the Eddington luminosity of the Sgr\,A$^{\star}$ Eddington supermassive black hole. 
Note that these results are exactly consistent with the independent estimate by \cite{tot06} as 
described above.

The above estimates are quite modest, and in our opinion, do not conflict with the existing 
observations. Yet, our arguments are not sufficiently strong to rule out the alternative conventional idea 
that the NPS originates from a nearby supernova remnant. Worth remarking in this respect is that the 
expansion velocity we derived conflicts with the order-of-magnitude
higher values advocated in the literature \citep{guo12a,guo12b,zub12,yan12,lac13}. 
We note that such large expansion velocities would result
in the formation of a very strong bow shock at the bubbles' boundaries, 
manifesting in very hot gas at the edges of the structure with temperatures of $kT \simeq 3-17$\,keV.
Currently there is no observational evidence for the presence of such a diffuse gaseous component 
around the Fermi Bubbles, or even within the Galaxy, except for its most central regions 
\citep[$\geq 5$\,keV plasma within $\sim 100$\,pc of Sgr\,A$^{\star}$, the origin of which
is still under debate;][]{koy07b}.

Finally, we note an interesting analogy between the Fermi Bubbles and the giant,
relic lobes in the nearby radio galaxy Centaurus\,A, which have been also resolved with
WMAP and \F \citep{har09, cena-lobe}, and for which the recent \SZ observations indicated 
analogously a rough pressure equilibrium with the surrounding medium \citep{sta13}. 
In both cases, the radiating ultra-relativistic electrons were proposed to be accelerated 
\emph{in-situ} via interaction with magnetic turbulence, 
rather than at weak shocks formed eventually at the bubbles/lobes edges. 
However, large-scale shock wave may anyway be required to generate turbulence accelerating 
high energy particles in the Fermi Bubbles 
\citep{mer11}. Reconnecting large-scale magnetic field may also play a role in this context 
in the Galactic Center region \citep{sof05}
as well as in the lobes of AGN \citep{gou12}. High-quality 
radio and X-ray observations of the Fermi Bubbles interiors, enabling a diagnosis of the plasma 
conditions and magnetic field structure similarly as in the case of the giant Centaurus\,A lobes
\citep{sul13,sta13} are needed to elaborate more on the particle acceleration processes 
at work (Cheung et al., in prep.).

\section{Conclusions}

In this paper we presented the results of our \SZ X-ray observations 
of high Galactic latitude regions positioned at the edges of the $\gamma$-ray Fermi Bubbles 
recently discovered with \F .
We showed that the detected diffuse X-ray emission is well reproduced by a
three-component plasma model including unabsorbed thermal emission of the Local Bubble
($kT \simeq 0.1$\,keV), absorbed thermal emission related to the North Polar Spur and/or 
Galactic halo ($kT \simeq 0.3$\,keV), and a power-law component at the level 
expected from the cosmic X-ray background. We did not find a non-thermal X-ray 
emission component associated with the bubbles exceeding $\sim$15$\%$ fluctuation 
of the CXB intensity, corresponding to 
$<9.3 \times 10^{-9}$\,erg\,cm$^{-2}$\,s$^{-1}$\,sr$^{-1}$ 
in the 2$-$10 keV energy range. Based on the gathered data, we argued 
that the North Polar Spur is possibly related to the bubbles rather than being a local phenomenon.
This followed from the indirect evidence we found for the 
presence of a large amount of  neutral matter absorbing the X-ray emission of 
the structure, as well as for a weak shock driven by the bubbles' expansion 
in the surrounding medium and compressing the halo gas to form the NPS feature.
In this scenario, we estimated the expansion velocity of the bubbles as 
$v_{\rm exp} \sim 300$\,km\,s$^{-1}$, corresponding to the shock Mach number 
$\mathcal{M} \simeq 1.5$.
We also showed that the non-thermal pressure 
and energy estimated by means of modeling the broad-band spectrum of the bubbles in the
framework of a simple one-zone leptonic model, $p_{\rm n/th}$ $\simeq$ 
$2 \times 10^{-12}$\,dyn\,cm$^{-2}$ and $E_{\rm n/th}$ $\simeq$ 
$10^{56}$\,erg, respectively, are in a rough equilibrium with the pressure 
and energy of the thermal plasma surrounding the bubble.

\acknowledgments 

We acknowledge the referee for a careful reading and for a number of useful and 
positive suggestions that helped improve the manuscript.
This work is partially supported by the Japanese Society for the 
Promotion of Science (JSPS) KAKENHI. We would like to thank 
Drs. M. Ishida and  N. Yamazaki for useful discussions on the 
nature of the GH plasma. 
\L .S. was supported by Polish NSC grant DEC-2012/04/A/ST9/00083.
Work by C.C.C. at NRL is supported in part by NASA DPR S-15633-Y.

\appendix
\section{X-ray spectra of compact sources}

As discussed in the text (Section 3.2.1), we believe that the bulk of the uncatalogued 
point-like X-ray features detected in our \SZ observations at the $\gtrsim 5 \sigma$ level are 
background AGN or galaxies and are unrelated to the GH or NPS structure. In this Appendix,
we first summarize the positions and the statistical significances of compact X-ray features 
detected above 3$\sigma$, and possible association of catalogued sources when possible (Tables 3 \& 4).
X-ray spectral fits of uncatalogued sources detected above $\gtrsim$ $5 \sigma$ level 
(with the assumed absorbed power-law model) are then summarized in Table 5.

\section{Abundance of the $kT \simeq 0.3$\,keV plasma component}

The limited photon statistics preclude us from precisely determining the metallicity
of the absorbed $kT \simeq 0.3$\,keV diffuse emission component, hence we
fixed $Z = 0.2 \, Z_{\odot}$ in the model fitting (see Section 3.2.2).
To validate our assumption, in Figure 9 we plot the temperature $kT$ versus 
the abundance for the N1$-$N8 pointings, derived when both $Z$ and $kT$ were allowed to vary.
The derived values cluster in a relatively narrow range of $kT \simeq 0.3$\,keV and 
$Z \simeq 0.2 \, Z_{\odot}$ indeed. We note the related discussion in \citet{wil03} and \citet{mil08}
concerning the brightest regions of the NPS targeted by \SZ and {\it XMM}, suggesting
the depleted C, O, Ne, Mg and Fe abundances of less than $0.5 \, Z_{\odot}$, but an 
enhanced N abundance which was however not clearly seen in our data.

\begin{figure}
\begin{center}
\includegraphics[angle=0,scale=0.7]{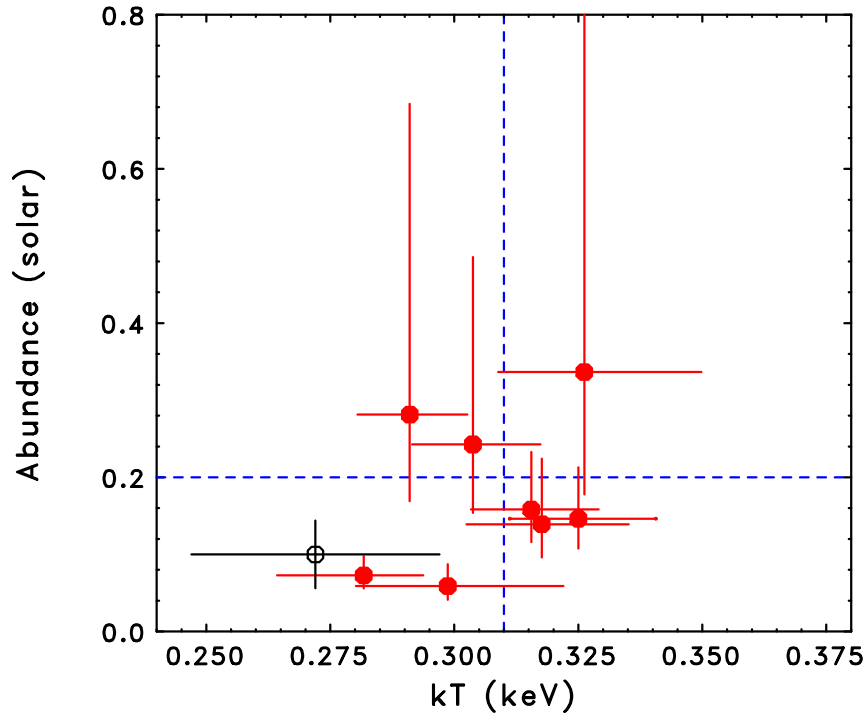}
\caption{Temperature $kT$ versus abundance $Z$ for the diffuse X-ray emission 
component \textsc{apec2} in the N1$-$N8 fields (filled red circles) and southern fields (average over S1$-$S6; open circle), derived assuming the three component 
plasma model introduced in Section 3.2.2, but with the abundance set free in the fitting
procedure. Blue dashed lines show the means of the best fit parameters, 
namely $kT = 0.31$\,keV and $Z = 0.2$\,$Z_{\odot}$.}
\end{center}
\end{figure}

\begin{figure}
\begin{center}
\includegraphics[angle=0,scale=0.7]{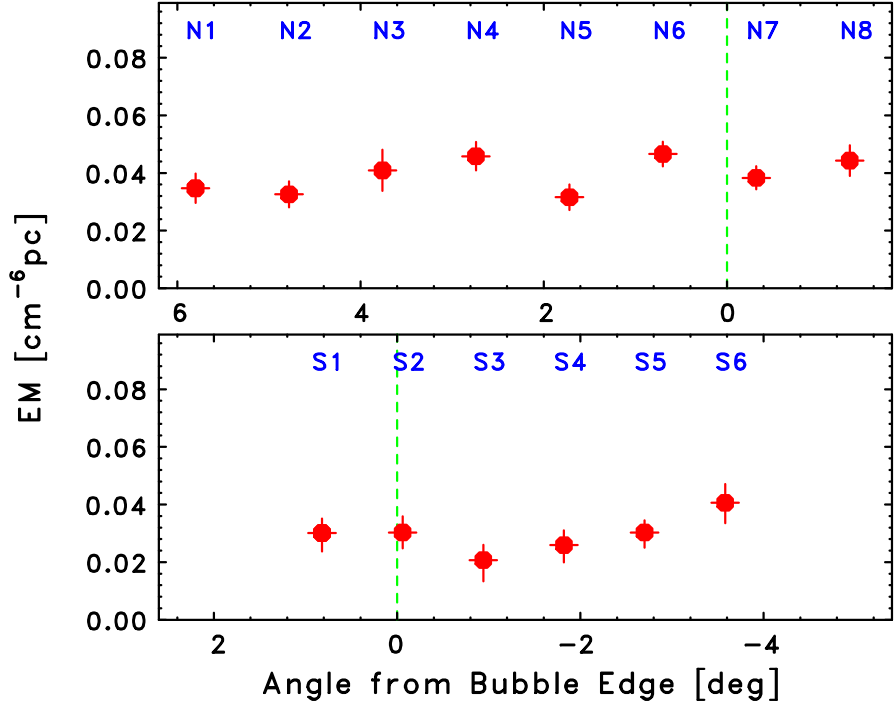}
\caption{
Variation in the spectral fitted EM parameter of the
\textsc{apec1} emission component in the north-east ({\it top}) and south ({\it bottom}) 
\SZ observations, as a function of the angular separation from the 
bubble edge. No significant changes in the EM at the expected position 
of the bubbles' boundary \citep[green dotted line following][]{su10} are seen, 
consistent with the expectation that the $kT \simeq 0.1$\,keV plasma component
is due to the Local Bubble emission contaminated by the solar wind charge exchange.} 
\end{center}
\end{figure}

\section{EM of $kT$ $\simeq$ 0.1 keV plasma}

Since the XIS is only sensitive to photon energies above 0.4\,keV, clearly distinguishing between the two 
diffuse thermal components ($kT \simeq 0.1$\,keV due to the LB/SWCX and 
$kT \simeq 0.3$\,keV due to the GH/NPS) is in general not easy. Therefore, it 
is in principle possible that a gradual jump visible in the derived EM between different 
neighboring fields (as seen in Figure 6), may be due to a sudden increase in the 
amount of contamination from the LB/SWCX component rather than a drop in the GH/NPS
component. In order to investigate this issue in more detail, in Figure 10 we plot the 
variation of the EM of the $kT = 0.1$\,keV plasma for the N1$-$N8 and S1$-$S6 pointings. 
As expected, no significant variations in the EM are observed,
thus confirming the reality of a gradual drop in the EM for the $kT \simeq 0.3$\,keV 
component claimed in Section 3.2.2.

\begin{deluxetable*}{cccccc}[th]
\tabletypesize{\scriptsize}
\tablecaption{List of compact X-ray features detected above 3$\sigma$ level and removed in the 
analysis of duffuse X-ray emission: Bubble North}
\tablewidth{0pt}
\tablehead{
\colhead{ID} &\colhead{R.A.} &  \colhead{DEC} & \colhead{stat sig$^a$} & \colhead{association$^b$} & \colhead{comment$^c$}\\
\colhead{} & \colhead{[$^{\circ}$]} & \colhead{[$^{\circ}$]} & ($\sigma$) & \colhead{} & \colhead{}
}
\startdata
src1  & 233.354 & 9.033 & 10.6 & $-$ &\\
src2  & 233.286 & 9.159 &  9.1 & $-$ &\\ 
      & 233.304 & 9.098 &  3.7 & $-$ &\\  
      & 233.503 & 9.015 &  4.5 & $-$ &\\ 
src3  & 233.692 & 8.032 &  4.9 & $-$ &\\
      & 233.623 & 8.015 &  4.5 & $-$ & low-E\\  
src4  & 233.825 & 7.110 &  4.9 & $-$ &\\
src5  & 233.933 & 7.175 &  6.2 & $-$ &\\
      & 233.870 & 7.108 &  3.8 & $-$ &\\  
      & 233.779 & 7.032 &  4.1 & $-$ &\\  
      & 233.710 & 7.057 &  4.5 & $-$ &\\  
      & 233.940 & 7.053 &  3.7 & $-$ & low-E\\  
      & 233.879 & 7.024 &  7.7 & Radio Source &\\  
      & 233.888 & 6.976 &  16.5 & QSO &\\  
      & 233.773 & 7.129 &  12.0 & QSO &\\  
      & 233.745 & 7.137 &  4.3 &  Radio Source &\\  
src6  & 234.102 & 6.160 &  6.7 &  $-$ &\\  
src7  & 234.001 & 6.120 &  5.7 &  $-$ &\\  
      & 234.100 & 6.101 &  3.3 & Radio Source &\\  
      & 234.018 & 6.150 &  4.4 & Galaxy &\\  
      & 233.982 & 6.145 &  4.0 & Galaxy &\\  
      & 234.000 & 6.195 &  3.5 & Galaxy &\\  
      & 234.262 & 5.151 &  3.7 & $-$ &\\  
      & 234.323 & 5.182 &  4.5 & $-$ &\\  
      & 234.285 & 5.191 &  4.1 & $-$ &\\  
src8  & 234.420 & 4.170 &  5.4 &  $-$ &\\  
src9  & 234.446 & 4.120 &  6.2 &  $-$ &\\  
      & 234.317 & 4.071 &  3.3 & $-$ & low-E\\  
      & 234.287 & 4.218 &  3.3 & $-$ & high-E\\
      & 234.378 & 4.147 &  7.8 & QSO &\\    
src10  & 234.481 & 3.210 &  5.8 &  $-$ &\\  
      & 234.478 & 3.252 &  3.3 & $-$ &\\  
src11  & 234.725 & 2.243 &  5.3 &  $-$ &\\  
      & 234.837 & 2.149 &  20.2 & QSO &\\    
      & 234.617 & 2.123 &  15.3 & QSO &\\    
\tableline
\enddata 
\tablecomments{$^a$: Statistical significance determined by the source detection algorithm in \textsc{ximage}. Only XIS0+3 data were used since the XIS1 has lower imaging quality due to higher instrumental background.\\
$^b$: Catalogue association of the source, if available.\\
$^c$: Sources detected only in the 0.5$-$2 keV energy range are denoted as ``low-E'', and sources detected only in the 2$-$10 keV energy range are denoted as ``high-E''.\\
}
\end{deluxetable*}

\begin{deluxetable*}{cccccc}[th]
\tabletypesize{\scriptsize}
\tablecaption{List of compact X-ray features detected above 3$\sigma$ level and removed in the 
analysis of duffuse X-ray emission: Bubble South}
\tablewidth{0pt}
\tablehead{
\colhead{ID} &\colhead{R.A.} &  \colhead{DEC} & \colhead{stat sig$^a$} & \colhead{association$^b$} & \colhead{comment$^c$}\\
\colhead{} & \colhead{[$^{\circ}$]} & \colhead{[$^{\circ}$]} & ($\sigma$) & \colhead{} & \colhead{}
}
\startdata
src1  & 332.660 & -45.883 & 4.8 & $-$ &\\
src2  & 332.705 & -45.902 & 5.8 & $-$ &\\
      & 332.649 & -45.758 &  13.7 & Star &\\    
      & 332.731 & -45.929 &  6.0 & Galaxy Group &\\    
      & 332.723 & -45.947 &  6.9 & QSO &\\
src3  & 331.432 & -45.662 & 4.8 & $-$ &\\    
      & 331.493 & -45.726 & 3.8 & $-$ &\\    
      & 331.509 & -45.664 & 3.8 & Star &\\
src4  & 330.338 & -45.613 & 8.3 & $-$ &\\    
src5  & 330.124 & -45.483 & 6.9 & $-$ &\\    
      & 330.225 & -45.423 & 4.0 & $-$ &\\    
      & 330.227 & -45.550 & 3.4 & $-$ &\\    
      & 330.169 & -45.395 & 5.3 & Radio Source &\\
src6  & 329.014 & -45.269 & 22.4 & $-$ &\\    
src7  & 329.155 & -45.310 & 7.1 & $-$ &\\    
      & 329.096 & -45.337 & 4.5 & Galaxy &\\
      & 329.041 & -45.507 & 10.6 & Radio Source &\\
src8  & 327.788 & -45.245 & 6.6 & $-$ &\\    
      & 327.863 & -45.330 & 3.6 & Star &\\
src9  & 326.635 & -45.203 & 4.8 & $-$ &\\    
src10  & 326.653 & -47.000 & 4.7 & $-$ &\\   
      & 326.688 & -45.191 & 3.4 & $-$ &\\   
      & 326.496 & -45.106 & 7.0 & Radio Source &\\  
\tableline
\enddata 
\tablecomments{As in Table~4, but for the sources in the south bubble observations.
}
\end{deluxetable*}

\begin{deluxetable*}{ccccc}[th]
\tabletypesize{\scriptsize}
\tablecaption{Fitting parameters for compact X-ray features detected above $\gtrsim$ 5$\sigma$ level}
\tablewidth{0pt}
\tablehead{
\colhead{ID} & \colhead{$N_{\rm H}$$^a$} & \colhead{PL index$^b$} & \colhead{PL flux$^c$} & \colhead{$\chi^2$/dof}\\
\colhead{}  & \colhead{($10^{20}$ cm$^{-2}$)} & \colhead{} & \colhead{(10$^{-14}$ erg cm$^{-2}$ s$^{-1}$)} & \colhead{}
}
\startdata
\multicolumn{5}{c}{Bubble North} \\
\tableline
src1  & 3.37(fix) & 2.44$_{-0.22}^{+0.23}$ & 6.19$^{+1.70}_{-1.46}$ & 1.29/41\\
src2   & 688$_{-340}^{+619}$ & 0.76$_{-0.60}^{+0.79}$ & 74.7$^{+21.2}_{-10.5}$ & 1.47/31\\
src3  & 3.83(fix) & 2.41$_{-0.46}^{+0.51}$ & 3.96$^{+2.55}_{-1.83}$ & 1.00/68\\
src4  & 3.86(fix) & 2.26$_{-0.38}^{+0.43}$ & 2.05$^{+1.02}_{-0.82}$ & 0.79/17\\
src5   & 3.86(fix) & 3.06$_{-0.76}^{+0.71}$ & 0.94$^{+1.30}_{-0.59}$ & 1.00/22\\
src6   & 4.06(fix) & 1.43$_{-0.33}^{+0.35}$ & 8.35$^{+2.97}_{-2.61}$ & 1.27/17\\
src7   & 4.06(fix) & 2.06$_{-0.30}^{+0.32}$ & 3.22$^{+1.20}_{-1.02}$ & 1.30/25\\
src8   & 4.45(fix) & 1.67$_{-0.37}^{+0.40}$ & 4.23$^{+1.70}_{-1.49}$ & 1.07/23\\
src9   & 4.45(fix) & 2.70$_{-0.45}^{+0.52}$ & 1.43$^{+0.95}_{-0.67}$ & 1.06/28\\
src10  & 4.26(fix) & 1.75$_{-0.30}^{+0.32}$ & 5.68$^{+1.95}_{-1.70}$ & 0.36/13\\
src11  & 5.02(fix) & 1.40$_{-0.41}^{+0.44}$ & 5.59$^{+2.56}_{-2.11}$ & 1.19/10\\
\tableline
\multicolumn{5}{c}{Bubble South} \\
\tableline
src1  & 1.84(fix) & 1.72$_{-0.41}^{+0.45}$ & 3.82$^{+1.90}_{-1.57}$ & 0.94/19\\
src2  & 1.84(fix) & 1.34$_{-0.51}^{+0.53}$ & 4.75$^{+2.49}_{-2.08}$ & 0.25/9\\
src3  & 1.66(fix) & 1.83$_{-0.49}^{+0.52}$ & 3.17$^{+1.91}_{-1.47}$ & 0.65/13\\
src4  & 1.89(fix) & 1.85$_{-0.33}^{+0.36}$ & 7.50$^{+3.08}_{-2.60}$ & 0.96/18\\
src5   & 1.89(fix) & 0.93$_{-0.29}^{+0.27}$ & 15.1$^{+3.7}_{-3.4}$ & 0.72/19\\
src6   & 2.16(fix) & 2.31$_{-0.13}^{+0.14}$ & 17.2$^{+3.1}_{-2.8}$ & 0.98/68\\
src7   & 2.16(fix) & 1.24$\pm$0.33 & 8.66$^{+2.79}_{-2.50}$ & 0.98/34\\
src8   & 2.45(fix) & 1.60$_{-0.27}^{+0.29}$ & 6.42$^{+2.07}_{-1.85}$ & 1.12/19\\
src9   & 3.03(fix) & 1.95$_{-0.65}^{+0.70}$ & 2.56$^{+2.03}_{-1.48}$ & 0.88/7\\
src10  & 3.03(fix) & 1.60$_{-0.48}^{+0.53}$ & 8.88$^{+5.80}_{-4.27}$ & 0.78/7\\
\tableline
\enddata 
\tablecomments{$^a$: Absorption column density fixed at the Galactic values, except for src2 in the North, which required an additional column density well exceeding $N_{\rm H,\,Gal}$.\\
$^b$: Spectral photon index in the single power-law model.\\
$^c$: Unabsorbed $2-10$\,keV flux.}
\end{deluxetable*}

\end{document}